\documentclass[twocolumns]{aa}
\usepackage{graphics,epsfig}

\begin{document}

\title{Dynamical and gravitational lensing properties of a new
phenomenological model of elliptical galaxies}

\author{C. Tortora \inst{1,2,4}, V.F. Cardone\inst{3}, E. Piedipalumbo\inst{1,4}}
\offprints{C.Tortora \email{ctortora@na.infn.it}} \institute{
Dipartimento di Scienze Fisiche, Universit\`{a} di Napoli Federico
II, Compl. Univ. Monte S. Angelo, 80126 Napoli, Italy \and
Osservatorio Astronomico di Capodimonte, Salita Moiariello, 16,
80131 - Napoli, Italy  \and Dipartimento di Fisica ``E.R.
Caianiello'', Universit{\`{a}} di Salerno, Via S. Allende, 84081 -
Baronissi (Salerno), Italy \and
 INFN - Sezione di Napoli}

\date{Received xxx /Accepted yyy}

\abstract{}{Recent observations of the line of sight velocity
profile of elliptical galaxies have furnished controversial
results with some works favouring the presence of a large amount
of dark matter in the outer regions and others arguing in favour
of no dark matter at all. In order to shed new light on this
controversy, we propose here a new phenomenological description of
the total mass profile of galaxies.}{Under the hypothesis of
spherical symmetry, we assume a double power\,-\,law expression
for the global mass\,-\,to\-\,light (hereafter $M/L$) ratio
$\Upsilon(r) = M(r)/L(r) = \Upsilon_0 (r/r_0)^{\alpha} (1 +
r/r_0)^{\beta}$ with $(\Upsilon_0, r_0)$ scaling parameters and
$(\alpha, \beta)$ determining the inner and outer slope of the
$M/L$ ratio. In particular, $\Upsilon \propto r^{\alpha}$ for
$r/r_0 << 1$ so that $\alpha = 0$ mimics a constant $M/L$ ratio in
the inner regions, while, for $(r/r_0 >> 1)$, $\Upsilon \propto
r^{\alpha + \beta}$ thus showing that models with $\alpha + \beta
= 0$ have an asymptotically constant $M/L$ ratio. A wide range of
possibilities is obtained by varying the slope parameters
$(\alpha, \beta)$ in the range we determine on the basis of
physical considerations. Choosing a general expression for the
luminosity density profile $j(r)$, we work out an
\textit{effective} galaxy model that accounts for all the
phenomenology observed in real elliptical galaxies. We derive the
main dynamics and gravitational lensing properties of such an
effective model.}{We analyze a general class of models, able to
take into account different dynamical trends. We are able to
obtain analytical expressions for the main dynamical and lensing
quantities. We show that constraining the values of $\alpha +
\beta$ makes it possible to analyze the problem of the dark matter
in elliptical galaxies. Indeed, positive values of $\alpha +
\beta$ would be a strong evidence for dark matter.}{Finally we
indicate possible future approaches in order to face the
observational data, in particular using velocity dispersion
profiles and lensed quasar events.}


\titlerunning{Dynamics and lensing in a phenomenological model for ellipticals}
\authorrunning{C. Tortora et al.}

\maketitle

\section{Introduction}

Applying the virial theorem to the Coma cluster to estimate its
mass, Zwicky (1937) provided the first evidence of the presence of
significant amounts of dark matter. Notwithstanding the great
theoretical, computational and observational efforts spent by both
astronomers and particle physicists, the nature, the properties
and the distribution of the elusive component remain still
unsolved problems more than 70 years after the pioneering work of
Zwicky. Solving these problems stands as one of the most
challenging yet fascinating issues of modern astrophysics.

Although detectable only through its gravitational effect, dark
matter leaves its imprint on a wide variety of observational
probes. Rotation curves of spiral galaxies (see, e.g., Sofue \&
Rubin 2001 for a comprehensive review) are a textbook example. The
presence of HI and H$\alpha$ gas makes it possible to reliably
determine the kinematics and dynamics of these systems. Modelling
the galaxy with the luminous and the gas components only is unable
to fit the observed rotation curve unless the
mass\,-\,to\,-\,light (hereafter $M/L$) ratio is unacceptably
higher than what is predicted on the basis of stellar population
synthesis models. Going up to the cosmological scales, the
remarkable success of the concordance $\Lambda$CDM model
(according to which the cosmological constant $\Lambda$ and the
cold dark matter are the main contributors to the total energy
density budget) in fitting almost the full set of available
astrophysical data (\cite{WMAP,Riess04}) provide a further strong
evidence in favor of dark matter. Moreover, in this same
framework, the presence of CDM is mandatory in order to reproduce
the observed matter power spectrum inferred from large galaxy
surveys (\cite{H03,P04}).

However, despite the great success on cosmological scales, the
concordance model is affected by serious shortcomings on the
galaxy scales. Indeed, numerical simulations of structure
formation in the $\Lambda$CDM scenario furnish some hints on the
mass profiles of dark matter haloes. While there is a general
consensus that relaxed haloes exhibit a density profile that is
well described by a double power law with outer asymptotic slope
$-3$, there is still controversy about the value of the inner
asymptotic slope with proposed values mainly in the range $\sim
1.0 - 1.5$, predicting the so called \emph{cuspy} mass density
models (see \cite{NFW96,NFW97,M98,Ghigna2000,P03,N03}). The first
results indicated an inner slope equal to $1$
(\cite{NFW96,NFW97}), on the contrary other studies generated a
more steeper slope equal to $1.5$ (\cite{M98,Ghigna2000,FM01}).
Whichever is the actual value of the inner slope, observations of
rotation curves strongly disfavor the presence of any cusp giving,
on the contrary, strong support to shallower density profiles (the
so called \emph{cored} models with an inner slope $<1$) in fierce
disagreement with results of simulations
(\cite{MdeB98,BE01,deB01}). It is worth stressing that the usual
modeling of spiral galaxies does not take into account the
presence of spiral arms and the possible clumpy distribution of
the gas component. Both these effects may introduce non circular
motions causing systematical errors in the determination of the
rotation curve or in inferring constraints on the mass model from
the fit to the data. Although recent higher resolution simulations
seem to partially alleviate both the cusp and the substructure
problems (\cite{M2006a,M2006b}), several alternative scenarios
have been proposed in order to solve these issues ranging from
modifications of the fundamental properties of dark matter
particles to changes in the laws of gravity (\cite{carroll}). A
different kind of approach is followed by M$\ddot{u}$cket \& Hoeft
(2003), that solving the Jean's equation with suitable assumptions
on the gravitational potential found an inner slope $ \leq 0.5$.
If confirmed, these contrasting results may offer the chance to
falsify the CDM paradigm on galaxy scales or, at least, can
furnish a fundamental support to drive N-body simulation analysis.

It is therefore interesting to look at elliptical galaxies in
order to avoid some of the observational problems described above
and to investigate whether dark matter is indeed ubiquitous as
expected. Unfortunately, here the situation is complicated by both
observational and theoretical difficulties. On one hand, the lack
of a well understood mass tracer makes it difficult to measure
kinematics out to the largest radii where dark matter is supposed
to dominate. On the other hand, interpreting the results is
complicated by the strong degeneracy between mass model and
anisotropy in the velocity space. As a consequence, the presence
of dark matter in elliptical galaxies is still a matter of debate.
Using different tracers and modeling techniques, Mould et al.
(1990) and Franx et al. (1994) found positive evidences of massive
dark matter halos around early type galaxies, while no dark matter
is needed to fit the measured velocity dispersion in some of the
galaxies considered by Bertin et al. (1994) and Gerhard et al.
(2001). Satellite dynamics (\cite{rom01}) and X\,-\,ray mass
estimates (\cite{LW99}) adds further support in favour of dark
matter, while opposite results have been recently obtained using
planetary nebulae to probe the outer regions
(\cite{Romanowsky03}). Furthermore, Mamon \& Lokas (2005a,b) have
recently shown that NFW\,-\,like density profiles are not able to
reproduce the global mass profile of elliptical galaxies whose
internal kinematics turn out to be well fitted by constant $M/L$
ratio models thus arguing against the presence of dark matter
haloes.

In order to improve our knowledge of dark matter haloes, one has
to look for alternative mass probes such as gravitational lensing.
Born as mere scientific curiosity predicted by general relativity,
gravitational lensing has nowadays given rise to a full sector of
modern astronomy given its versatility that renders it an ideal
tool to investigate a wide range of astrophysical phenomena (see,
e.g., \cite{SEF} and \cite{PLW01} for illuminating textbooks).
Since elliptical galaxies represent $\sim 80\%$ of lens galaxies
in multiply imaged quasar systems (\cite{FT91}), reconstructing
the mass model for these systems gives interesting constraints on
the dark matter haloes surrounding the lens (see, e.g.,
\cite{K95,RK97,KKF98,TK04} for a far to be exhaustive list of
references). Galaxy\,-\,galaxy lensing (\cite{Fisher00,GS02}) is
another opportunity to investigate dark matter resorting to the
weak lensing regime.

While the use of lensing data alleviates the problem with
observational issue, there is still a theoretical prejudice
concerning the choice of dark matter profile, which can crucially
influence the results. Even if there is a long and illustrious
tradition of mass models for early type galaxies, ranging from
distribution function based ones, to mass density profile or
gravitational potential potential based ones, we present a new
parametrization of the mass profile, describing the M/L ratios as
a double power law. Thus, we propose here to replace the actual
elliptical galaxy with an {\it effective} one with a mass profile
$M(r)$ given as $\Upsilon(r) {\times} L(r)$, with $L(r)$ the
deprojected luminosity distribution. A key role is played by the
global $M/L$ ratio $\Upsilon(r)$ which we parameterize as
$\Upsilon(r) = \Upsilon_{0} (r/r_0)^{\alpha} (1 + r/r_0)^{\beta}$
with $\Upsilon_0$ a strength parameter, $r_0$ a length scale,
while $\alpha$ and $\beta$ determines the asymptotic behaviors of
the $M/L$ ratio for $r/r_0 << 1$ and $r/r_0 >> 1$, respectively.
The ansatz adopted for $\Upsilon(r)$ has the virtue of being quite
simple so that many dynamical and lensing quantities may be
expressed analytically or in terms of special functions.
Nevertheless, it is also quite versatile giving the possibility to
mimic the main features of a wide range of models\footnote{Another
possible phenomenological approach has been recently investigated
by (\cite{Nap04}), but they only considered an empirical gradient
of the $M/L$ ratio assuming an a priori model for the dark matter
halo density profile.}. For instance, constant $M/L$ models are
obtained by setting $(\alpha, \beta) = (0, 0)$, while models with
$\alpha + \beta > 0$ mimic the effect of dark matter haloes on the
global $M/L$ ratio. Moreover, in the inner regions the main
kinematic and dynamical quantities does not depend on $\beta$, and
if we set $\alpha = 0$ will only depend on the $M/L$ strength
$\Upsilon_0$ and the parameters which characterize the luminosity
profile in agreement with the tentative conclusion of Mamon \&
Lokas (\cite{MamonLokas05a}) that mass follows light in the
interiors of elliptical galaxies.

The paper is organized as follows. In Sect. 2, we introduce our
parametrization of the global $M/L$ ratio discussing the role
played by the slope parameters $(\alpha, \beta)$ and the
asymptotic behaviours. The basic properties of the resulting
effective galaxy model are presented in Sect. 3 where we also
compare our model with some widely used alternative descriptions
of elliptical galaxies. Dynamical and lensing properties of the
model are presented with great detail in Sects. 4 and 5
respectively. A summary of the main results and some perspectives
on constraining the model parameters through observations are
presented in the concluding Sect. 6.

\section{A variable M/L ratio}\label{sec:mass_to_light}

The approach to the problem of dark matter in elliptical galaxies
is a two step process, where, first, the projected light
distribution is deconvolved to a three dimensional mass density by
assuming a constant $M/L$ ratio and second, a dark halo mass
profile is added to the luminous component.The theoretically
predicted luminosity weighted velocity dispersion (projected along
the line of sight) is then compared to the observational data, in
order to determine the stellar $M/L$ ratio and the parameters of
the halo model. A successful result of such a fitting procedure is
considered as an evidence of the presence of dark matter, while
the possibility of fitting the same data with a constant $M/L$
model and no dark halo argues against the presence of this unseen
component. Needless to say, the procedure crucially depends on the
choice of the model for the dark matter distribution.

In order to alleviate such a problem we work out a family of
galaxy models which is able to reproduce, through an appropriate
parametrization, a wide variety of behaviours in the galaxy mass
density distribution, including both the situation where the dark
matter is dominating as well the situation where the stellar
component is predominant As a consequence such a kind of models
can be used to analyze and shape a whole sample of galaxies in a
homogeneous way. To this end, it is worth noting that in constant
$M/L$ models, the mass follows the light, while this does not hold
anymore in presence of a dark halo. Actually, defining
$\Upsilon(r) \equiv M(r)/L(r)$ with $M(r)$ and $L(r)$ the total
mass and luminosity respectively within a distance $r$ from the
galaxy center, it is straightforward to understand that, in
presence of a dark matter halo, $M(r)$ still increases for values
of $r$ beyond the visible edges of the galaxy so that
$\Upsilon(r)$ turns out to be monotonically increasing function in
the outer regions. On the opposite, in absence of a dark matter
halo, $M(r)$ follows $L(r)$ thus leading to $\Upsilon(r) \sim {\rm
constant}$. Note, however, that, strictly speaking, having a
constant $M/L$ is not a sufficient probe against the presence of
dark matter. Actually, it is possible that $\Upsilon(r)$ is a weak
function of $r$ over the range probed by the data so that
detecting its variation is quite hard if the data are not of very
good quality. Should, however, $\Upsilon(r) \simeq \Upsilon_0
\simeq \Upsilon_{\star}$ with $\Upsilon_{\star}$ the stellar $M/L$
ratio predicted by population synthesis models, then one could
indeed argue that there is no need of dark matter. As a
consequence, we may reconsider the problem of dark matter in
elliptical galaxies as that of directly understanding whether
$\Upsilon(r)$ is constant or not and, if yes, what is its trend as
a function of the radius. Thus, we assume a phenomenological
ansatz for the $M/L$ ratio that is able to smoothly interpolate
between the two opposite cases. The only other ingredient needed
to model the galaxy is then the luminosity density. Although it is
in principle possible that the stellar $M/L$ ratio changes with
$r$ because of a distance dependent distribution of massive stars,
it is worth stressing that color gradients are often too low to
motivate a strong variation of $\Upsilon_{\star}$. As a
consequence, we may safely consider an increasing $M/L$ as a clear
signal of a dark matter halo.

Assuming spherical symmetry, we propose the following expression
for the $M/L$ ratio\,:

\begin{equation}
\Upsilon(r) \equiv \frac{M(r)}{L(r)} = \Upsilon_0 \left (
\frac{r}{r_0} \right )^{\alpha} \left ( 1 + \frac{r}{r_0} \right
)^{\beta} \label{eq: mtol}
\end{equation}
with $\Upsilon_0$ a scaling $M/L$ ratio, $r_0$ a reference radius
and $(\alpha, \beta)$ two slope parameters that we will constrain
later on the basis of physical considerations only. Note that for
$(\alpha, \beta) = (0, 0)$ the model reduces to the constant $M/L$
case with $\Upsilon(r) = \Upsilon_0$. A no dark matter model is
then obtained setting $(\alpha, \beta) = (0, 0)$ and $\Upsilon_0
\simeq \Upsilon_{\star}$.

The slope parameters $(\alpha, \beta)$ drive the asymptotic
behaviors of $\Upsilon(r)$ \,:

\begin{equation}
\Upsilon(r) \propto \left \{
\begin{array}{ll}
r^{\alpha} & {\rm for} \ \ \ r/r_0 << 1 \\
~ & ~ \\
r^{\alpha + \beta} & {\rm for} \ \ \ r/r_0 >> 1 \\
\end{array}
\right . \label{eq: asym}
\end{equation}
so that it is clear that $\alpha$ determines the slope of the
$M/L$ ratio in the inner regions, while $\beta$ enters the
determination of the outer slope. As shown in Fig.\,1, the
reference radius $r_0$ marks the transition from the inner to the
outer asymptotic slopes.

\begin{figure}
\centering \resizebox{8cm}{!}{\includegraphics{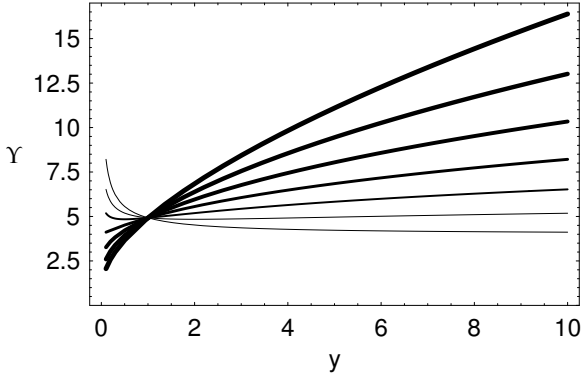}}
\caption{Global $M/L$ ratio $\Upsilon$ as a function of $y \equiv
r / r_{0}$ setting $\Upsilon_{0} = 4$, $\beta=0.3$. The inner
slope $\alpha$ takes values from $-0.3$ to $0.3$ (in steps of 0.1)
from the lightest to the boldest curve.}\label{fig:MtoL_vs_y}
\end{figure}

Eq. (\ref{eq: asym}) helps drawing some useful qualitative
considerations to constrain the parameters $(\alpha, \beta)$.
First, let us remind that, since both $M(r)$ and $L(r)$ vanish at
$r = 0$, we expect that $\Upsilon(r)$ could diverge, take a
constant value or converge to $0$. In Fig. \ref{fig:MtoL_vs_y} we
plot $\Upsilon$ as a function of the radial coordinate $r$, fixing
$\beta$ and allowing $\alpha$ to vary and assume both negative and
positive values. An inspection of the figure shows us that
negative values of $\alpha$ give diverging values of $\Upsilon$
for $r \to 0$, on the contrary null or positive $\alpha$ give a
$\Upsilon$ that becomes finite or null. Having $\Upsilon(r)
\rightarrow \infty$ as $r \rightarrow 0$ could seem unrealistic
since one could argue for a huge content of dark matter in the
inner regions, but this is not the case, since it is possible that
$L(r) \to 0$ more rapidly than $M(r)$.

On the other hand, it is easy to see that models with $\alpha +
\beta = 0$ describes galaxies with finite total mass since, in
this case, both $M(r)$ and $L(r)$ for $r >> r_0$ are constants.
Moreover, since $M(r)$ is asymptotically increasing or flat, while
$L(r)$ is constant, $\Upsilon(r)$ should be an increasing or flat
function of $r$ for $r >> r_0$ thus leading to the constraint
$\alpha + \beta \ge 0$. As we will see later, we will recover
these same constraints from a different and more physically
motivated perspective.

As a final remark, let us remember that we have assumed spherical
symmetry. Actually, real galaxies are hardly spherically
symmetric, but are intrinsically ellipsoidal or triaxial systems.
However, an ellipsoidal or triaxial version of the model we will
investigate may be obtained by replacing in Eq. (\ref{eq: mtol})
the spherical radius $r$ with the elliptical $m = (x^2 + y^2 +
z^2/q^2)^{1/2}$ or the triaxial one $\tilde{m} = (x^2 + y^2/p^2 +
z^2/q^2)^{1/2}$ with $q$ and $p$ the major\,-\,to\,-\,minor and
intermediate\,-\,to\,-\,minor axial ratios (\cite{BT87}). Models
which result from equation (\ref{eq: mtol}), after such
replacement, are similar to the spherical ones in some important
properties such as  mass profiles. Moreover, it is worth
remembering the existence of a degeneracy among mass, anisotropy
and flattening that makes it possible, at least in principle, to
mimic a flattening of the system with an anisotropic velocity
distribution (\cite{MB01}). Finally, using spherical symmetry
allows us to express the main quantities of interest in terms of
analytical or special functions.

\section{Basic properties}\label{sec:basics}

The ansatz (\ref{eq: mtol}) is the main ingredient of our approach
to the dark matter problem in elliptical galaxies. In order to
complete the description of a galaxy, we need to choose a
luminosity density profile that, when projected on the plane of
the sky, is able to fit the observed surface brightness. The ideal
route to follow should be to deproject the Sersic profile
(\cite{Sersic}) since this is known to well fit the surface
brightness of elliptical galaxies (\cite{CCD93,GC97,PS97}).
Although this is mathematically possible (\cite{MC02}), the result
is expressed in terms of unusual special functions and is
therefore of little practical utility. A different way is to
choose a luminosity density profile whose two dimensional
projection mimics the de Vaucouleurs law (\cite{deV48}) since this
profile also gives a good fit to elliptical galaxies. A versatile
family of models suitable for our study is provided by the Dehnen
(1993) models whose luminosity density is\,:

\begin{equation}
j(r) = j_0 \left ( \frac{r}{a} \right )^{-\gamma} \left ( 1 +
\frac{r}{a} \right )^{\gamma - 4} \label{eq: jr}
\end{equation}
with $j_0$ and $a$ scaling quantities and $\gamma$ a positive
slope parameter. Note that the model reduces to the Jaffe (1983)
and Hernquist (1990) ones for $\gamma = 2$ and $\gamma = 1$
respectively. The luminosity density scales as $r^{-\gamma}$ at
the centre, while asymptotically decreases as $r^{-4}$. The
constraint $\gamma < 3$ ensures that the model is physically well
behaved, but, actually, only models with $\gamma \le 2$ are
interesting since their surface density fits well the de
Vaucouleurs law provided the scale radius $a$ is expressed in
terms of the effective radius $R_e$. In particular, for $\gamma =
1$, it is $a = 1.33 (1 + \sqrt{2})^{-1} R_e$, while similar
relations hold for other values of $\gamma$ (\cite{Dehnen93}).
Models with $2 < \gamma < 3$ are still meaningful, but have some
disturbing features such as an infinite central potential and a
diverging central velocity dispersion. In the following, we will
set $\gamma$ as a free parameter, but in the figures we will set
$\gamma = 1$ in order to recover the widely used Hernquist model.

Being the model spherically symmetric, it is immediate to evaluate
the luminosity density. Inserting Eq. (\ref{eq: jr}) into the
standard relation

\begin{displaymath}
L(r) = \int_{0}^{r}{4 \pi r'^2 j(r') dr'}
\end{displaymath}
we easily get\,:

\begin{equation}
L(r) = L_{tot} x^{3 - \gamma} (1 + x)^{\gamma - 3} \label{eq: lum}
\end{equation}
with $x \equiv r/a$ and $L_{tot}$ the total luminosity given by\,:

\begin{equation}
L_{tot} = \frac{4 \pi j_0 a^3}{3 - \gamma} \ . \label{eq: ltot}
\end{equation}
Note that the constraint $\gamma < 3$ ensures that the total
luminosity is finite and that $L(r)$ does not diverge at the
centre as it is physically reasonable.

\subsection{Mass profile}

Having chosen a model for the luminosity density, the mass profile
of the galaxy is immediately obtained from the definition of $M/L$
ratio and the ansatz (\ref{eq: mtol})\,:

\begin{equation}
M(x) = \Upsilon_0 \ L_{tot} \ x_0^{-\alpha} \ x^{3 + \alpha -
\gamma} \ (1 + x)^{\gamma - 3} \ \left ( 1 + \frac{x}{x_0} \right
)^{\beta} \label{eq: mass}
\end{equation}
with $x_0 \equiv r_0/a$. In particular, we note that, the larger
are $\beta$ or $\alpha$, the greater is the value of the total
mass. For $\alpha + \beta =0$ we have a finite mass.

Eq. (\ref{eq: mass}) makes it possible to put physically motivated
constraints on the slope parameters $(\alpha, \beta)$. First, the
mass profile must vanish at the centre so that we get\,:

\begin{equation}
M(r = 0) = 0 \rightarrow 3 + \alpha - \gamma > 0 \iff \alpha >
\gamma - 3 \ . \label{eq: massvinc}
\end{equation}
Let us now consider the asymptotic behaviour at
infinity\footnote{Although a finite total mass is a desirable
feature, there are plenty of interesting models with a formally
infinite total mass that are typically truncated {\it by hand} at
the virial radius thus obtaining a finite total mass.}\,:

\begin{equation}
\lim_{r \rightarrow \infty}{M(r)} = \left \{
\begin{array}{ll}
\infty & {\rm for} \ \ \ \alpha + \beta > 0 \\
\Upsilon_0 \ L_{tot} & {\rm for} \ \ \ \alpha + \beta = 0 \ \ \ . \\
0 & {\rm for} \ \ \ \alpha + \beta < 0 \\
\end{array}
\right . \label{eq: totmass}
\end{equation}
Needless to say, a null total mass is unphysical so that
Eq.(\ref{eq: totmass}) leads to the constraint\,:

\begin{equation}
\alpha + \beta \ge 0 \iff \alpha \ge - \beta \ . \label{eq:
totmassvinc}
\end{equation}
Note that the case $\alpha + \beta = 0$ individuates models with
finite total mass as we have argued above directly from Eq.
(\ref{eq: mtol}).

\subsection{Mass density profile}

For spherically symmetric models, the mass density may be
evaluated differentiating the mass profile. Inserting Eq.(\ref{eq:
mass}) into the usual relation

\begin{displaymath}
\rho(r) = \frac{1}{4 \pi r^2} \frac{dM}{dr} = \frac{1}{4 \pi a^3
x^2} \frac{dM}{dx}
\end{displaymath}
we easily get\,:

\begin{equation}
\rho(r) = \frac{\Upsilon_0 \ L_{tot}}{4 \pi a^3 x_{0}^{\alpha +
1}} \ x^{\alpha - \gamma} \ (1 + x)^{\gamma - 4} \ \left ( 1 +
\frac{x}{x_0} \right )^{\beta - 1} {\cal{P}}(x) \label{eq: rho}
\end{equation}
with

\begin{equation}
{\cal{P}} \equiv (\alpha + \beta) x^2 + \left [ 3 + \alpha (1 +
x_0) + \beta - \gamma \right ] x + (3 + \alpha - \gamma) x_0 \ .
\label{eq: defp}
\end{equation}
The mass density in Eq.(\ref{eq: rho}) must be physically
reasonable so that we can further constrain $(\alpha, \beta)$. To
this aim, let us consider its asymptotic behaviours. In the inner
regions, it is\,:

\begin{equation}
\rho(x) \propto x^{\alpha - \gamma} \ \ \ {\rm for} \ \ \ x << 1 \
. \label{eq: asymrhoinn}
\end{equation}
First, we note that models with a cusp are obtained for $\alpha -
\gamma < 0$, while a flat inner core is present for models with
$\alpha - \gamma = 0$. Moreover, since the mass density cannot
increase with $r$, we get the constraint\,:

\begin{equation}
\alpha - \gamma \le 0 \rightarrow \alpha \le \gamma \ . \label{eq:
vincinnrho}
\end{equation}

Let us study now the behaviour in the opposite limit, i.e. for $x
>> 1$. In this case, we have to consider separately models with
infinite or finite total mass. In the former case, $\alpha + \beta
\ne 0$ and we get\,:

\begin{equation}
\rho(x) \propto x^{\alpha + \beta - 3} \ \ \  {\rm for} \ x >>1 \
, \label{eq: asymrhoinf}
\end{equation}
while for models with $\alpha + \beta = 0$, it is\,:

\begin{equation}
\rho(x) \propto \left \{
\begin{array}{ll}
x^{-4} & {\rm for} \ \ \ 3 - \gamma \ne \alpha x_0 \\
x^{-5} & {\rm for} \ \ \ 3 - \gamma = \alpha x_0 \\
\end{array}
\right . \label{eq: asymrho}
\end{equation}
Imposing that $\rho$ is a decreasing function at infinity makes it
possible to further constrain the slope parameters. Considering
Eqs.(\ref{eq: asymrhoinf}), we thus get\,:

\begin{equation}
\alpha + \beta - 3 < 0 \rightarrow \alpha < 3 - \beta \ .
\label{eq: vinoutrho}
\end{equation}

\subsection{Surface mass density}

It is worth deriving the surface mass density of the model since
this will enter the derivation of the lensing quantities which
Sect.\,5 is devoted to. To this aim, we start by the usual
definition\,:

\begin{equation}
\Sigma = \int_{-\infty}^{+\infty} \rho(r_{1}, r_{2}, r_{3})
dr_{3}, \label{eq:surf_dens}
\end{equation}
with $(r_1, r_2)$ cartesian coordinates in the plane orthogonal to
the line of sight and $r_3$ along the line of sight. Inserting the
mass density (\ref{eq: asymrho}) in Eq.(\ref{eq:surf_dens}) and
changing to cylindrical coordinates, one gets (see Appendix A in
\cite{Sand04})\,:

\begin{displaymath}
\Sigma(\xi) =  \frac{L_{tot}\,\Upsilon_{0}\, x_{0}^{-\alpha -
\beta}}{2\,\pi \,{r_{s}}^2} \, \int_{0}^{\pi /2}  \xi^{1 + \alpha
- \gamma} \, \csc(\theta)^{2 + \alpha  - \gamma}
\end{displaymath}
\begin{displaymath}
\,\,\,\,\,\,\,\,\,\, {\times} \, ( 1 + \xi \,\csc(\theta ) )^{-4 +
\gamma}\, (x_{0} + \xi \,\csc(\theta ) ) ^{-1 + \beta}\,
\end{displaymath}
\begin{displaymath}
\,\,\,\,\,\,\,\,\,\, {\times} \, [x_{0}\, (3 + \alpha  - \gamma  ) +
( 3 + \alpha  + x_{0}\, \alpha + \beta  - \gamma  ) \,\xi \,
\csc(\theta)
\end{displaymath}
\begin{equation}
\,\,\,\,\,\,\,\,\,\, + \left( \alpha  + \beta  \right) \,{\xi
}^2\, \csc (\theta )^2 \, ] \, d \theta. \label{surdens}
\end{equation}
where $\xi = R/a$, with $R = \sqrt{r_{1}^{2}+r_{2}^{2}}$.
Eq.(\ref{surdens}) may be analytically evaluated only in the case
$x_0 = 1$ otherwise numerical integration is needed. However, also
for $x_0 = 1$, the result turns out to be a rather messy
combination of special functions so that it is preferable to
resort to numerical integration also in this case.

The asymptotic trends of the surface mass density depend on which
is the range $\alpha$ lies in. If $\gamma - 2 < \alpha < \gamma
-1$\footnote{See Eq. (\ref{eq: vincinnvc}) to motivate the lower
bound, more stringent than Eq. (\ref{eq: massvinc})}, $\Sigma(\xi)
\rightarrow \xi^{\alpha - \gamma + 1}$ for $\xi \to 0$, while,
$\Sigma$ asymptotes to a constant value in the same limit for
$(\gamma - 1) \leq \alpha \le \gamma$. These regimes were
qualitatively analyzed in the previous section and can be
inspected in Fig. \ref{fig:MtoL_vs_y}, where it is clear how the
inner trend of the $M/L$ ratio is crucially dependent on the
choice of $\alpha$. On the other hand, $\Sigma \propto \xi^{\alpha
+ \beta -2}$ for $\xi \rightarrow \infty$, which is expected given
the outer slope of the three dimensional mass density.

Further constraints on the slope parameters $(\alpha, \beta)$ may
be obtained considering the above asymptotic scaling of
$\Sigma(\xi)$. Indeed, imposing that $\Sigma(\xi)$ is always
decreasing, $(\alpha, \beta)$ must satisfy the constraints $\alpha
- \gamma +1 < 0$ and $\alpha + \beta - 2 < 0$. In particular, for
$\gamma = 1$, the first constraint translates to $\alpha < 0$, and
an asymptotic flat $\Sigma(\xi)$ for $\xi \to 0$ is obtained when
is $\alpha \geq 0$. Thus, $\alpha$ crucially influences the inner
trend of $\Sigma$ similarly to what done for $\Upsilon$. The other
constraint further reduces the range of variability of $\alpha$
and $\beta$ in Eq. (\ref{eq: vinoutrho}), giving the condition
$\alpha < 2 - \beta$.

\subsection{A general comment on our effective model}

A general comment is in order here. The philosophy motivating our
approach to the dark matter in elliptical galaxies is favored by
the phenomenology of the problem. We have adopted an ansatz for
the $M/L$ ratio and a model for the luminosity distribution and
are now working out the basic properties of a system that may be
described by these two assumptions. Strictly speaking, this system
does not represent the actual galaxy, but it is an {\it effective}
model of the galaxy. Should dark matter be indeed present, the
galaxy must be correctly modelled as a two components system with
a dark halo embedding the luminous component. However, from the
point of view of an observer, the dynamical quantities that can be
measured and compared with the theoretical predictions (such as
the rotation curve and the velocity dispersion) are determined by
the total mass and density profile only. As a consequence,
describing the galaxy with an effective model is
phenomenologically equivalent to modeling the system with two
different components. Actually, one could fit an effective galaxy
model and then decompose it in a luminous component plus a dark
matter halo determining the density profile of the halo as $\rho_h
= \rho - \Upsilon_{\star} j(r)$ with $\rho$ given by Eq.(\ref{eq:
rho}). In the phenomenological approach we are pursuing here, we
only need to study the dynamical and lensing properties of the
effective model defined by Eqs.(\ref{eq: mtol}) and (\ref{eq:
jr}), while disentangling the two components is not necessary.

It is, nevertheless, interesting to compare the relative
contributions of $\rho_h$ and $\rho_{\star} = \Upsilon_{\star}
j(r)$ to the total mass density $\rho$. Depending on the values of
the parameters $(\alpha, \beta, x_0)$, the two components may
furnish comparable contributes to $\rho$ for medium radii ($x$
around $1$ or few units), while, as expected, it is the dark
matter halo to dominate the outer mass budget. Nevertheless, given
the way we have assigned the model, the dark matter density has
the same asymptotic behaviours of the total one, i.e. $\rho_h
\propto x^{\alpha - \gamma}$ for $x \rightarrow 0$ and $\rho_h
\propto x^{\alpha + \beta - 3}$ for $x \rightarrow \infty$, thus
the inner slope of the dark matter density is determined by the
luminous component through the value of $\gamma$. As a
consequence, the constraints on the model parameters $\alpha$ and
$\gamma$ provide information about the competition between the
luminous and dark matter of galaxies, especially in the inner
regions where the baryons are supposed to be dominating
(\cite{MamonLokas05a}).

\subsection{Local M/L ratio}

A general comment is in order here. Eq. (\ref{eq: mtol}) defines
the $M/L$ ratio as $M(r)/L(r)$. This is also referred to in
literature as {\it global} $M/L$ ratio, while the quantity
$\Upsilon_{loc}(r) = \rho(r)/j(r)$, with $\rho(r)$ and $j(r)$ the
mass and luminosity density, is dubbed {\it local} $M/L$ ratio.

Now it is interesting to compute the local $M/L$ ratio as\,:

\begin{equation}
\Upsilon_{loc}(x) \equiv \frac{\rho(x)}{j(x)} = \frac{\Upsilon(x)
{\cal{P}}(x)}{(3 - \gamma) x_0 (1 + x/x_0)} \ . \label{eq:
locmtol}
\end{equation}
Apart the constant numerical factor (that is positive definite for
$\gamma < 3$), the local $M/L$ ratio is similar to the global
$M/L$ ratio, but is no more a double power law because of the
multiplicative term ${\cal{P}}(x)/(1 + x/x_0)$. Note that, in the
inner regions $(x << 1)$, both $\Upsilon(x)$ and
$\Upsilon_{loc}(x)$ scale as $x^{\alpha}$. On the other hand, the
behaviour at infinity is radically different. To this end, let us
note that, for $x >> 1$, $\Upsilon_{loc}$ scales as $x^{\alpha +
\beta + 1}$, whereas Eq.(\ref{eq: asym}) shows that $\Upsilon
\propto x^{\alpha + \beta}$. As a consequence, models with finite
total mass (i.e. $\alpha + \beta = 0$) have an asymptotically
constant global $M/L$ ratio, while a divergent local $M/L$ ratio.
We thus argue in favour of using $\Upsilon$ rather
$\Upsilon_{loc}$ in order to avoid any disturbing divergence.

\subsection{Comparison with a generalized NFW profile}

It is interesting to compare our phenomenological ansatz for the
$M/L$ ratio with that coming out from an assumed dark halo model.
To this aim, describing in both cases the luminosity component
with a Dehnen model, we have to choice a dark matter density
profile among the various proposals available in literature.
Motivated by the results of a wide set of numerical N\,-\,body
simulations, it is customary to model dark halos through the so
called NFW model whose mass density profile reads (Navarro et al.
1996,1997)\,:

\begin{equation}
\rho_{NFW}(r)=\rho_{s}\bigg (\frac{r}{r_h}\bigg )^{-1}\bigg
(1+\frac{r}{r_h}\bigg )^{-2},
\end{equation}
where $r_h$ and $\rho_s$ scaling values for the radius and the
density. Following the common practice, it is more convenient to
parametrize the model by the virial mass $M_v$ and the
concentration $c = r_v/r_s$ with $r_v$ the virial radius. As it is
easily seen, $\rho_{NFW}$ asymptotically scales as $r^{-1}$ and
$r^{-3}$ in the inner and outer regions respectively. While there
is a wide consensus on the value of the outer slope, a fierce
debate is still open on the inner scaling with values as high as
1.5 (\cite{M98,Ghigna2000,FM01}). As a consequence, it is more
instructive to consider the generalized NFW (gNFW) model defined
as\,:

\begin{equation}
\rho_{gNFW}(r)=\rho_{s}\bigg (\frac{r}{r_h}\bigg )^{-\delta}\bigg
(1+\frac{r}{r_h}\bigg )^{\delta - 3},
\end{equation}
which reduces to the NFW model for $\delta = 1$, while $\delta = 1.5$
reproduces the results obtained by Moore et al. (1998). Note that the total
mass is formally infinite, but, as usual, we truncate the model at the
virial radius and take $M_v$ as finite total mass.

In the rest of this section, we will compare the $M/L$ ratio and the mass
density profile of our effective galaxy with the same properties of a real
galaxy modelled with a Dehnen\,-\,like luminous component and a gNFW halo.
For illustrative purposes, we also consider the case of a single gNFW
component although this leads to less reliable results in the inner baryon
dominated regions.

\subsubsection{$M/L$ ratio comparison}

The global $M/L$ ratio for the gNFW\,+\,Dehnen (hereafter gNFW+D) models
may be easily computed as $\Upsilon_{gNFW+D} = [M_{gNFW} + \Upsilon_{\star}
L(r)]/L(r)$, while for the gNFW only model it is simply $\Upsilon_{gNFW} =
M_{gNFW}/L(r)$.

Depending on the values of $(\gamma, \delta)$, we can have different
variation of $\Upsilon_{gNFW+D}$ with $r$. For instance, if $\delta >
\gamma$, $\Upsilon_{gNFW+D}$ presents a mimimum and increases outwards, while
in the inner region may eventually diverge. Similar trends may be observed
also for $\Upsilon_{gNFW}$ which is not surprising given that
$\Upsilon_{gNFW+D}(r) = \Upsilon_{\star} + \Upsilon_{gNFW}(r)$. Our
phenomenological ansatz for $\Upsilon(r)$ in Eq.(\ref{eq: mtol}) presents
similar characteristics with $\alpha$ playing the role of $\delta$. In
particular, models with $\alpha < 0$ presents a minimum and may be
divergent at the centre.

The similar role played by $\alpha$ and $\delta$ may be well
understood considering the asymptotic trends of
$\Upsilon_{gNFW+D}$ and $\Upsilon_{gNFW}$. First, when $\gamma <
\delta$, for $r \rightarrow 0$ both $\Upsilon_{gNFW+D}$ and
$\Upsilon_{gNFW}$ scales as $r^{\gamma - \delta}$ so that both
quantities diverge. On the other hand, if $\gamma \ge \delta$, we
get $\Upsilon_{gNFW+D}(0) = \Upsilon_{\star}$, while
$\Upsilon_{gNFW}(0) = 0$. In the opposite limit, for $r
\rightarrow \infty$, we get $\Upsilon_{gNFW+D} = \Upsilon_{gNFW} =
M_{tot}/L_{tot}$, i.e. an asymptotically constant value.

Our ansatz for $\Upsilon$ scales as $r^{\alpha}$ in the inner
regions and as $r^{\alpha + \beta}$ in the outer ones, being
therefore able to recover the trends of gNFW+D and gNFW models.
For instance, in the case of gNFW+D model, the inner slope is
recovered for $\alpha = \gamma - \delta$ or $\alpha = 0$ depending
on the $\gamma$ being smaller or larger than $\delta$, while an
asymptotically constant value is obtained for $\alpha + \beta = 0$
 \footnote{We note that the gNFW
model, also being able to reproduce observed flat rotation curve,
has a total finite mass and therefore an asymptotically flat $M/L$
ratio. However, this is not a serious shortcoming such a profile,
since only the behaviour in a limited and finite region is
important to describe the observations and on these scales the
model reproduces a nearly flat rotation curve.}. Our model can
reproduce asymptotically both flat $M/L$ (for $\alpha + \beta =0$)
and increasing $M/L$ (for $\alpha + \beta =1$) giving more
possibilities respect to the gNFW.

\subsubsection{Mass density comparison}

For what concerns the mass density, the NFW+Dehnen model presents
the following trends

\begin{equation}
\rho_{gNFW+D}(r) \propto \left \{
\begin{array}{ll}
r^{-\delta} & {\rm for} \ \ \ r \to 0   {\rm \ \ and \ \ \delta \geq \gamma} \\
r^{-\gamma} & {\rm for} \ \ \ r \to 0   {\rm \ \ and \ \ \delta \leq \gamma}  \\
~ & ~ \\
r^{-3} & {\rm for} \ \ \ r \to \infty   \\
\end{array}
\right .
\end{equation}
The gNFW model has the same value for the outer slope, while
$\rho_{gNFW} \propto r^{-\delta}$ for $r \rightarrow 0$.

Since the mass density of our effective model scales as $r^{\alpha
- \gamma}$ for $r \rightarrow 0$, we are able to recover the same
inner asymptotic trends of the gNFW+D and gNFW model by suitably
choosing $\alpha$. Considering the case $\gamma = 1$, the choice
$\alpha = 0$ gives $\rho \propto r^{-1}$ as for the NFW model
($\delta = 1$), while for $\alpha = -0.5$ we get the Moore model
($\delta = 1.5$). Positive values of $\alpha$ do not recover the
trend of gNFW+D model, since in this case $\rho \propto
r^{-\epsilon}$ with $\epsilon \in [0,1)$, and $\rho_{gNFW+D}
\propto r^{-1}$. In the case of gNFW model, assuming $\gamma=1$,
the models with $\delta > 1$ ($<1$) are recovered when $\alpha$ is
negative (positive). Finally, we note that, once $\alpha$ has been
set in order to recover the same inner asymptotic trend of a given
gNFW+D model, the outer $r^{-3}$ scaling may be recovered by
setting $\beta = -\alpha$ since our mass density scales as
$r^{\alpha + \beta - 3}$ in these regions (as stated above when we
analyzed the M/L ratio).

\section{Dynamical quantities}\label{dynamics}

After having evaluated the main properties of the mass model, we
determine here some useful dynamical quantities. In a first step,
we derive the gravitational potential discussing its main
properties. From an observational point of view, however, the
rotation curve and the velocity dispersion are more appealing
quantities since they can be compared to the observed data in
order to investigate the viability of our parametrization of the
$M/$L ratio.

\begin{figure*}
\centering \resizebox{14cm}{!}{\includegraphics{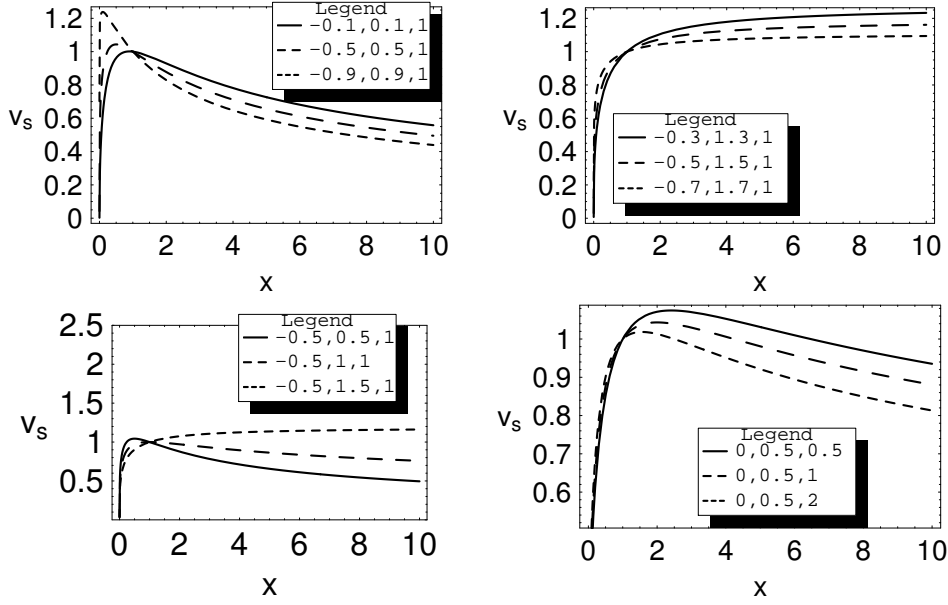}}
\caption{Normalized circular velocity $v_{s} \equiv
\frac{v_{c}(x)}{v_{c}(1)}$ for different combinations of the
parameters $(\alpha,\beta,x_{0})$. Starting on the top, in the
panel on the left we consider model with finite mass ($\alpha +
\beta =0$) and on the right models with flat curves ($\alpha +
\beta =1$). In the bottom, on the left we fix $\alpha$ and change
$\beta=0.5,1.0,1.5$ and on the right $\alpha=0$, $\beta=0.5$ and
we change $x_{0}=0.5,1,2$.} \label{fig:velocity_vs_x}
\end{figure*}

\subsection{Gravitational potential}

Although not explicitly needed to evaluate the dynamical quantities of
interest, it is nonetheless useful to compute the gravitational potential
for the model we are considering. For spherical systems, it is
(\cite{BT87})\,:

\begin{displaymath}
\Phi(r) = - \frac{G M(r)}{r} - 4 \pi G \int_{r}^{\infty}{\rho(r') r' dr'}
\end{displaymath}
that, inserting Eqs.(\ref{eq: mass}) and (\ref{eq: rho}) gives\,:

\begin{equation}
\Phi(x) = - \frac{G \Upsilon_0 \ L_{tot}}{a \ x_0^{\alpha}} \left
[ \frac{x^{3 + \alpha - \gamma}}{(1 + x)^{3 - \gamma}} \left ( 1 +
\frac{x}{x_0} \right )^{\beta} + \frac{{\cal{F}}(x; {\bf p})}{x_0}
\right ] \label{eq: phi}
\end{equation}
with ${\bf p}$ the set of parameters $(\alpha, \beta, \gamma, x_0)$ and \,:

\begin{equation}
{\cal{F}}(x; {\bf p}) \equiv \int_{0}^{\infty}{\xi^{\alpha -
\gamma + 1} (1 + \xi)^{\gamma - 4} (1 + \xi/x_0)^{\beta - 1}
{\cal{P}}(\xi) d\xi} \ . \label{eq: defeffe}
\end{equation}
Note that this expression only holds for models with $\alpha +
\beta \leq 1$ and is a complicated combination of hypergeometric
and Appell functions. As an example, we report the result for $x_0
= 1$ which is the simplest one\,:

\begin{equation}
{\cal{F}}(x) = \left [ \frac{\gamma - 3 - \alpha}{\alpha + \beta -
2} {\cal{F}}_1(x) + \frac{(\alpha + \beta) x}{\alpha + \beta - 1}
{\cal{F}}_2(x) \right ] x^{\alpha + \beta - 2} \label{eq: effe}
\end{equation}
having defined\,:

\begin{equation}
{\cal{F}}_1(x) \equiv {_2F_1(2 - \alpha - \beta, 4 - \beta -
\gamma, 3 - \alpha - \beta, -1/x)} \ , \label{eq: defeffeuno}
\end{equation}

\begin{equation}
{\cal{F}}_2(x) \equiv {_2F_1(1 - \alpha - \beta, 4 - \beta -
\gamma, 2 - \alpha - \beta, -1/x)} \ , \label{eq: defeffedue}
\end{equation}
with ${_2F_1(a, b, c, y)}$ the hypergeometric function\footnote{We use here
the same notation for the hypergeometric function as in the {\it
Mathematica} package.} (\cite{GR80}).

The case $\alpha + \beta = 1$ merits some discussion since the function
${\cal{F}}$ turns out to be constant. Since the gravitational potential is
defined up to an arbitrary additive constant, we may choose this constant
in such a way that $\Phi(x)$ for this class of models simplifies to $ -
GM(x)/x$ which is the well known Keplerian potential of a spherical system
with finite total mass.

Finally, we study the asymptotic trends. For $x \to \infty$, $\phi
\propto x^{-1 + \alpha + \beta}$, thus in order to have a
plausible gravitational potential, the models with $\alpha + \beta
> 1$ are ruled out, so that we conclude that the only physical
models are those with
\begin{equation}
\alpha + \beta - 1 \leq 0  \rightarrow \alpha \leq 1- \beta.
\label{eq: constr_grav_pot}
\end{equation}

\subsection{Circular velocity}

For a spherically symmetric mass distribution the circular velocity is
simply given by
\begin{equation}
v_{c}=\sqrt{\frac{G M(r)}{r}}
\end{equation}
which, in our case, becomes\,:

\begin{equation}
v_c(x) = \sqrt{\frac{G \Upsilon_0 \ L_{tot}}{a \ x_0^{\alpha}} \ x^{2 +
\alpha - \gamma} \ (1 + x)^{\gamma - 3} \ (1 + x/x_0)^{\beta}} \ . \label{eq:
vc}
\end{equation}
It is interesting to look at the asymptotic trends\,:

\begin{equation}
v_c^2(x) \propto \left \{
\begin{array}{ll}
x^{2 + \alpha - \gamma} & {\rm for} \ \ \ x << 1 \\ ~ & ~ \\ x^{\alpha +
\beta - 1} & {\rm for} \ \ \ x >> 1 \\
\end{array}
\right . \ . \label{eq: asymvc}
\end{equation}
First, we note that $v_c$ must vanish at the centre so that we get the
condition\,:

\begin{equation}
2 + \alpha - \gamma > 0 \rightarrow \alpha > \gamma - 2 \label{eq:
vincinnvc}
\end{equation}
that is more stringent than Eq.(\ref{eq: massvinc}). We consider
as physically motivated those models having an asymptotically flat
or decreasing rotation curve. Thus, using Eq. (\ref{eq: asymvc}),
we obtain the constraint in Eq. (\ref{eq: constr_grav_pot}).

In particular, models with $\alpha + \beta = 0$ presents a
keplerian falloff of the rotation curve consistent with the finite
total mass, while $v_c$ is asymptotically flat when $\alpha +
\beta = 1$ with\,:

\begin{equation}
v_c(\infty, \alpha + \beta = 1) = \sqrt{\frac{G \Upsilon_0 \
L_{tot}}{a x_0}} \ . \label{eq: vcinfty}
\end{equation}

The qualitative dependence of $v_c$ on the parameters
$(\alpha,\beta, x_0)$ may be derived by
Fig.\,\ref{fig:velocity_vs_x} where we plot the scaled circular
velocity $v_s \equiv v_c(x)/v_c(x = 1)$ for some representative
choices of the parameters $(\alpha, \beta, x_0)$ and setting
$\gamma = 1$. In the top left panel, we present models having
finite total mass (i.e., with $\alpha + \beta = 0$) thus showing a
keplerian falloff of the rotation curve. The top right panel,
instead, refers to models with asymptotically flat rotation curve
that is with $\alpha + \beta - 1 = 0$ corresponding to a $M/L$
ratio that is still increasing beyond the visible edge of the
galaxy as in presence of a dark matter halo. In both cases, for a
given $x$, $v_c$ is a decreasing or increasing function of $\beta$
depending on $x$ being larger or lower than 1 respectively. The
situation is reversed when considering the dependence on $\alpha$
with the higher values of $\alpha$ leading to larger (smaller)
circular velocities in the regime $x > 1$ ($x < 1$). The role of
$\beta$ is further investigated in the bottom left panel where we
hold $\alpha$ fixed and change $\beta$ showing that this slope
parameter determines the transition from asymptotically decreasing
to asymptotically flat rotation curves. In particular, models with
flat rotation curves rotate faster than those with a keplerian
falloff. Finally, in the bottom right panel, we investigate the
dependence on the scaling radius $x_0$ finding out that the larger
is $x_0$, the lower is $v_c(x)$. It is worth noting that the wide
variety of cases is an evidence of the extreme versatility of the
adopted parametrization even if this originates degeneracies among
the parameters $(\alpha, \beta, x_0)$ that may not be disentangled
with observations of the rotation curve only.

\begin{figure}
\centering \resizebox{8cm}{!}{\includegraphics{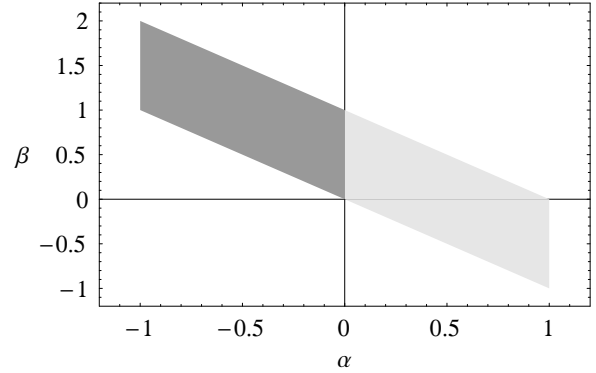}}
\caption{Constraints on model parameters $\alpha$ and $\beta$ for
$\gamma = 1$. The selected region is allowed by a theoretical
analysis on the model and its main related quantities (mass, mass
density, surface mass density, gravitational potential and
circular velocity). The parameters $\alpha$ and $\gamma$ are
crucial in this analysis, in particular to determine the low radii
trend of the surface mass density. In the limit $r \rightarrow 0$,
the darker region corresponds to diverging (`cuspy') $\Sigma$,
while the lighter one individuates models with flat (`cored')
$\Sigma$.}\label{fig:constr}
\end{figure}

Combining all the constraints obtained insofar on model
parameters, we end up with the following range for $\alpha$\,:
\begin{equation}
\max{(\gamma - 2, -\beta)} \le \alpha \le \min{(\gamma, 1 -
\beta)} \  \label{eq: rangealpha}
\end{equation}
which we have graphically summarized in Fig.\,\ref{fig:constr} for
models with $\gamma = 1$.

\subsection{Velocity dispersion}

While the circular velocity is related to the ordered motions in
the galactic plane, the velocity dispersion takes into account
disordered motions and thus is the main quantity to describe
dynamics of non rotating systems like elliptical galaxies.
Assuming an isotropic velocity dispersion tensor, the solution of
the Jeans equation gives the following general formula
(\cite{BT87})\,:

\begin{equation}
\sigma^2(r) = \frac{1}{j(r)} \int_{r}^{\infty}{j(r') \frac{G
M(r')}{r'^2} dr'} \label{eq:sigma_def}
\end{equation}
that, applied to our particular case, reduces to\,:

\begin{equation}
\sigma^2(x) = \frac{G \Upsilon_0 \ L_{tot}}{a \ x_0^{\alpha}}
{\times} \frac{I_{\sigma}(x, {\bf p})}{x^{\alpha - \gamma} \ (1 +
x)^{\gamma - 4} (1 + x/x_0)^{\beta - 1} {\cal{P}}(x)} \label{eq:
sigma}
\end{equation}
with\,:

\begin{equation}
I_{\sigma} \equiv  \int_{x}^{\infty} {\xi^{1 + 2 (\alpha -
\gamma)} (1 + \xi)^{2 \gamma - 7} (1 + \xi/x_0)^{2 \beta - 1}
{\cal{P}}(\xi) d\xi} \ . \label{eq: defisig}
\end{equation}
This integral is only defined for $\alpha + \beta < 2$ which is
not a problem given that $\alpha$ nd $\beta$ ranges in the
interval defined by Eq.(\ref{eq: rangealpha}). The general
analytical expression is again a complicated combination of
hypergeometric and Appell functions so that a numerical
integration turns out to be easier to handle in the applications.
As an example, we only report the case with $x_0 = 1$ that is\,:

\begin{displaymath}
I_{\sigma}(x) = \frac{x^{2 (\alpha + \beta) - 5}}{2 (\alpha +
\beta - 2) \left [ 2 (\alpha + \beta) - 5 \right ]} {\times}
\end{displaymath}

\begin{displaymath}
\ \ \ \ \ \ \ \ \left \{ \ 2 (2 - \alpha - \beta) (3 + \alpha -
\gamma) \ {\cal{I}}_1 \ + \right .
\end{displaymath}

\begin{equation}
\ \ \ \ \ \ \ \  \left . \left [ \ \alpha (2 \alpha + 4 \beta - 5)
+ \beta (2 \beta - 5) \right ] \ x \ {\cal{I}}_2 \ \right \}
\label{eq: isig}
\end{equation}
where we have set\,:

\begin{equation}
{\cal{I}}_1 \equiv {_2F_1(5 - 2 \alpha - 2 \beta, 7 - 2 \beta - 2
\gamma, 6 - 2 \alpha - 2 \beta, -1/x)} \ , \label{eq: defisiguno}
\end{equation}

\begin{equation}
{\cal{I}}_2 \equiv {_2F_1(4 - 2 \alpha - 2 \beta, 7 - 2 \beta - 2
\gamma, 5 - 2 \alpha - 2 \beta, -1/x)} \ . \label{eq: defisigdue}
\end{equation}
A wide variety of cases is obtained varying the parameters $(\alpha, \beta,
x_0)$. Note, however, that what is indeed observed is not the velocity
dispersion $\sigma(x)$, but rather its luminosity weighted projection along
the line of sight that we will compute later.

\subsubsection{Effects of anisotropy}

Eq.(\ref{eq:sigma_def}) only holds under the hypothesis that the
velocity dispersion is the same along the three axes of the
coordinate system. In the case of a spherically symmetric system,
the relation $\sigma_{\theta} = \sigma_{\phi} = \sigma_t$ holds,
where $\sigma_{\theta}$, $\sigma_{\phi}$ and $\sigma_{t}$ are
respectively the polar, azimuthal and tangential velocity
dispersion and we can define the anisotropy parameter:

\begin{figure}
\centering \resizebox{8.5cm}{!}{\includegraphics{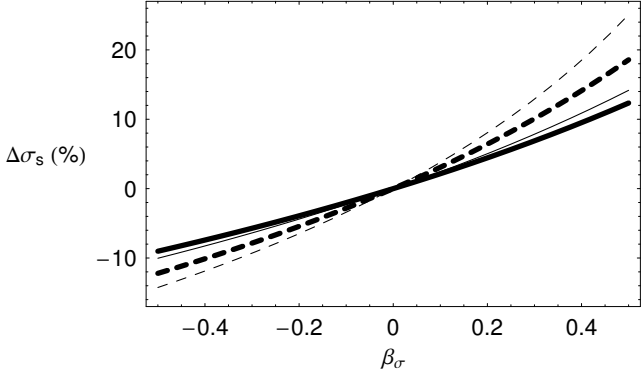}}
\caption{$\Delta \sigma_{s}$ as a function of $\beta_{\sigma}$ for
$x = 0.1$ (thin dashed), $x = 0.5$ (thick dashed), $x = 2$ (thin
solid) and $x = 10$ (thick solid). The model parameters are set as
$(\alpha, \beta, \gamma, x_0) = (-0.5, 1.0, 1.0, 1.0)$.}
\label{fig:plot_anis_const}
\end{figure}

\begin{equation}
\beta_{\sigma}(r) \equiv 1 -
\frac{\sigma_{t}^{2}}{\sigma_{r}^{2}},
\end{equation}
with $\sigma_r$ the radial velocity dispersion; isotropy in the
velocity space means $\beta_{\sigma} = 0$ (i.e., when
$\sigma_{r}=\sigma_{t}$). In the general case,
Eq.(\ref{eq:sigma_def}) becomes\,:

\begin{equation}
\sigma^2(r) = \frac{1}{\eta (r) j(r)} \int_{r}^{\infty}{\eta(r')
j(r') \frac{G M(r')}{r'^2} dr'},
\end{equation}
where $\eta(r)$ is obtained by solving\,:

\begin{equation}
\frac{d \log \eta}{d \log r}= 2 \beta_{\sigma}(r).
\end{equation}

To quantify the effect of the anisotropy, we first consider the
simplest model $\beta_{\sigma}(r) = const$ and introduce the
\emph{anisotropy to isotropy ratio parameter} $\Delta \sigma_{s} =
100 \times [\sigma_s(x, \beta_\sigma)/\sigma_s(x, 0) - 1]$,
defined as a function of scaled velocity dispersions
\footnote{$\sigma_{s}(x) \equiv \frac{\sigma(x)}{\sigma(1)}$, is a
scaled version of the velocity dispersion.}. In
Fig.\,\ref{fig:plot_anis_const} we plot $\Delta \sigma_{s}$ as a
function of $\beta_{\sigma}$ for different values of the
dimensionless radius $x$. It turns out that a constant anisotropy
significantly alters the velocity dispersion profile, with
positive values increasing the velocity dispersion at low and high
radii. The amount of the relative variation depends on $x$ and the
model parameters $(\alpha, \beta, \gamma, x_0)$.

\begin{figure}
\centering \resizebox{8.5cm}{!}{\includegraphics{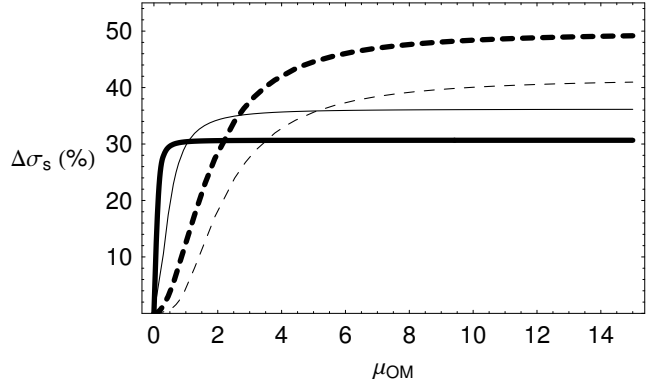}}
\caption{$\Delta \sigma_{s}$ as a function of $\mu_{OM}$. The
values of $x$, the line coding and the model parameters are the
same as in Fig.\,\ref{fig:plot_anis_const}.}
\label{fig:plot_anis_OM}
\end{figure}

A constant anisotropy, although simple and widely used, is not a
realistic assumption. More realistic results are obtained
considering the Osipkov\,-\,Merrit parametrization
(\cite{Osipkov79,Merritt85})\,:

\begin{equation}
\beta_{\sigma}(r)=\frac{r^{2}}{r^{2}+r^{2}_{OM}}
\end{equation}
with $r_{OM}$ the anisotropy radius. Isotropic models are
recovered in the limit $r_{OM} \to \infty$. In such a case we can
extend the definition of anisotropy to isotropy ratio parameter as
$\Delta \sigma_s = 100 \times [\sigma_s(x, x_{OM})/\sigma_s(x,
\infty) - 1]$, where $x_{OM}\equiv r_{OM}/a$. In
Fig.\,\ref{fig:plot_anis_OM} we plot $\Delta \sigma_s$ as a
function of $\mu_{OM}\equiv \frac{1}{x_{OM}}$. Although the
results are more complicated, it is nevertheless clear that also
in this case the anisotropy introduces large changes in the
velocity dispersion. $\Delta \sigma_{s}$ converges to $0$ for high
$x_{OM}$ (low $\mu_{OM}$) and assumes saturated and very high
values for low $x_{OM}$ (high $\mu_{OM}$).

\subsubsection{Projected velocity dispersion}

\begin{figure*}
\centering \resizebox{15cm}{!}{\includegraphics{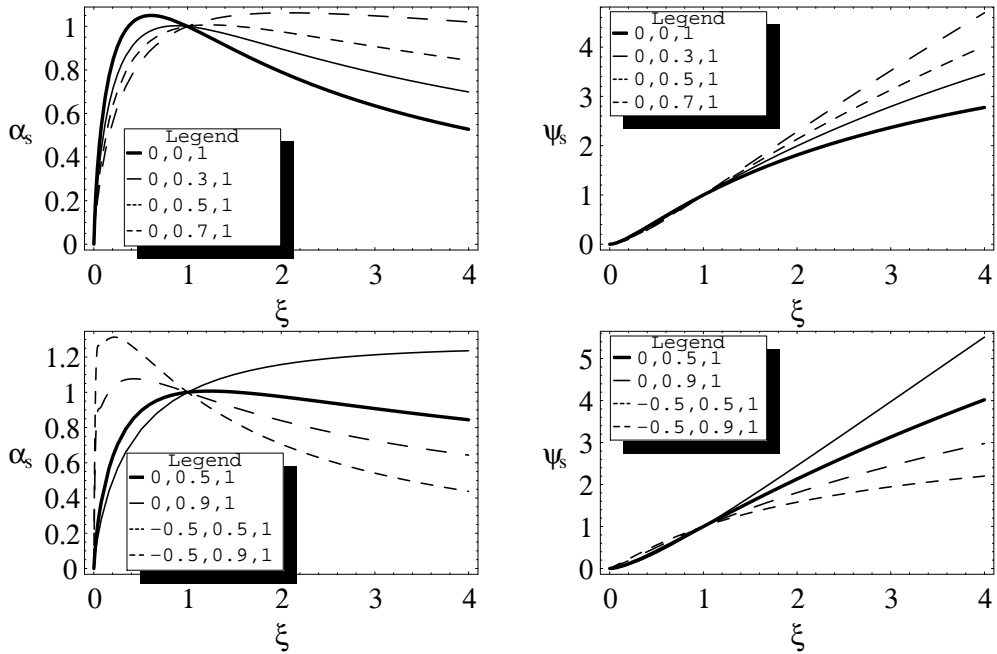}}
\caption{Normalized deflection angle $\alpha_{s}$ (left) and
lensing potential $\psi_{s}$ (right) as function of $\xi$. The
model parameters $(\alpha, \beta, x_0)$ are set as in the inset of
each panel, while $\gamma = 1$ is used for the luminosity
profile.} \label{def_angle}
\end{figure*}

The velocity dispersion defined by Eq.(\ref{eq:sigma_def}) is not
directly measurable. Indeed, when comparing to observations, we
have to take into account some effects. First, the line of sight
does not coincide with the radial direction so that we measure
only the projection of the velocity dispersion along the line of
sight. Moreover, all the galaxy elements contribute to the
measured $\sigma$, but the more luminous is the element, the
larger will be its contribution to the observed quantity. As such,
we have first to define the velocity dispersion projected along
the line of sight and luminosity weighted. Denoting by
$\sigma_{los}$ this quantity, it is (\cite{BT87})\,:

\begin{equation}
\sigma_{los}^2 = \frac{2}{I(R)} \int_{R}^{\infty} j(r) \frac{G
M(r) \sqrt{r^2 - R^2}}{r^2} dr \label{eq: sigmalos}
\end{equation}
with $I(R)$ the luminosity intensity.

As a second step, we have to take into account the seeing that makes the
intrinsic intensity differs from the observed one. We therefore define the
seeing corrected velocity dispersion as\,:

\begin{equation}
\sigma_{seeing}^{2}(R)=\frac{\int d^{2}R' P(R-R') I(R')
\sigma_{p}^{2}(R')}{\int d^{2}R' P(R-R')I(R')},
\end{equation}
where $P(R-R')$ is the point spread function taking into accont both the
atmospheric and the instrument seeing. Finally, we have to consider the
effect of the spatial binning and the finite width of the slit used to make
the measurements. Therefore, the measured quantity reads\,:

\begin{equation}
\sigma_{bin}^{2}=\frac{\int_{A}{dA'I_{s}(R') \sigma^{2}_{seeing}(R')}}{\int_{A}{dA'I_{s}(R')}},
\end{equation}
where $A$ is the area of the slit and $I_{s}$ is the seeing
corrected intensity. We do not investigate these observational
effects since they are strictly related to the observational
setup. We, however, stress the need to carefully take into account
these corrective terms in order to not introduce systematic errors
when comparing the model to the data.

\section{Gravitational lensing}\label{sec:lensing}

Considered at its early beginning a little more than a scientific
curiosity, gravitational lensing has now given rise to what may be
referred to as {\it lensing astronomy}. Being directly dependent
on the mass rather than the light distribution, gravitational
lensing has turn out to be an invaluable tool to investigate the
total density profile in early type galaxies at intermediate and
high redshift. The lensing observable quantities (such as the
position of the images and their flux ratios) in multiply imaged
sources (quasars or galaxies) make it possible to constrain the
mass profile. Combining lensing reconstruction with a measurement
of the velocity dispersion helps breaking degeneracies inherent in
lens modeling thus shedding a powerful light on their dark matter
content (\cite{TK02,TK04}). Understanding the mass distribution of
the lens plays a key role also in the determination of the Hubble
constant $H_0$ through the time delay method (\cite{Refsdal}). In
particular, Kochanek (2002) has shown that the predicted delay
between two images, and hence the derived value of $H_0$, is
primarily governed by the average surface mass density in the
annulus defined by their radial distances from the lens center
with a small correction taking into account the slope of the
profile in this annulus. A careful determination of mass profile
is therefore mandatory to get a reliable estimate of $H_0$ and
hence of the distance scale. Lens modeling plays a vital role also
in cosmological applications of gravitational lensing. Since
early\,-\,type galaxies are expected to represent at least 80$\%$
of lenses (\cite{TOG84,FT91}), a correct description of their
luminous and dark component is an essential ingredient in every
attempt to constrain cosmological parameters through statistics of
lens systems.

Motivated by these considerations, we complement our presentation of our
new phenomenological model by a detailed investigation of its lensing
properties.

\subsection{Deflection angle and lensing potential}

For a spherically symmetric model, the deflection angle $\hat{\alpha}$ may
be easily computed as\,:

\begin{equation}
\hat{\alpha}(\xi) = \frac{2 a}{\xi}\int _{0}^{\xi}{\kappa(u) u du}
= \hat{\alpha}_1 {\times} \alpha_{s}(\xi; {\bf p})
\end{equation}
where the normalized deflection angle $\alpha_{s}$ depends on the
dimensionless radius $\xi = R/a$ and the model parameters ${\bf p} =
(\alpha, \beta, \gamma, x_0)$, while $\hat{\alpha}_1$ is the value of the
deflection angle for $\xi = 1$. The normalized surface mass density
$\kappa$ is defined as the ratio $\Sigma/\Sigma_{crit}$ between the surface
mass density $\Sigma$ and the critical surface density $\Sigma_{crit}
\equiv c^{2}D_{s}/(4 \pi G D_{d}D_{ds})$ with $D_{s}$, $D_{d}$ and $D_{ds}$
the source, lens and lens\,-\,source angular diameter distances\footnote{In
the rest of the paper we adopt the flat concordance $\Lambda$CDM model with
$(\Omega_M, \Omega_{\Lambda}, h) = (0.3, 0.7, 0.72)$ consistent with the
WMAP measurements (\cite{WMAP}).}.

\begin{figure*}
\centering \resizebox{17cm}{!}{\includegraphics{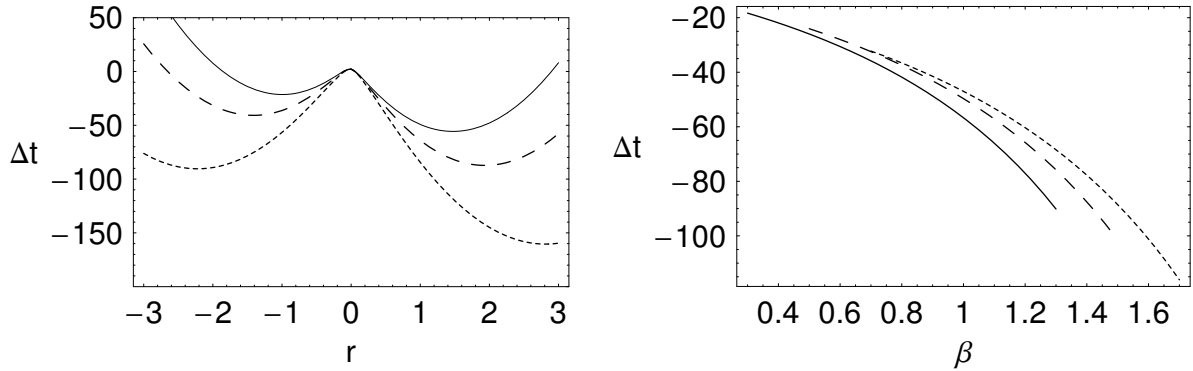}}
\caption{{\it Left.} Time delay function $\Delta t$ (measured in
days) as a function of the radial distance from the centre
(measured in ${\rm arcsec}$). Note that negative values of $r$
corresponds to radial distances after a rotation of $\pi$ radians.
Model parameters are set as $(\alpha, \Upsilon_0, x_0) = (-0.3,
5.0, 1.0)$ and three values of $\beta$ are considered, namely
$\beta = 0.8$ (solid), $\beta = 1.0$ (long dashed), $\beta = 1.2$
(short dashed). For the luminosity profile, we set $\gamma = 1$,
while the source position is $r_s = 0.3 \ {\rm arcsec}$. The
extremals of the function corresponds to the images. {\it Right.}
$\Delta t$ as function of $\beta$ for three values of $\alpha$,
namely $\alpha = -0.3$ (solid), $\alpha = -0.5$ (long dashed) and
$\alpha = -0.7$ (short dashed). The remaining model parameters are
set as before, while we move the source at $r_s = 0.1 \ {\rm
arcsec}$ and evaluate $\Delta t$ at $r = 1 \ {\rm arcsec}$.}
\label{Dt_all}
\end{figure*}

Although quite complicated analytical expression are possible for
$\alpha_s$, we prefer to not report them here and discuss the results
looking at the left panels in Fig.\,\ref{def_angle}. In particular, in the
top left panel, we consider models with an inner core ($\alpha = 0$) and
find out that the larger is $\beta$, the higher (the lower) is the
deflection angle for $\xi > 1$ ($\xi < 1$). Note that, in the outer
regions, $\alpha_s(\xi)$ is always larger than what is predicted by a
constant $M/L$ model. In the bottom left panel, we show the effect of
changing $\alpha$ while holding $\beta$ fixed. It turns out that increasing
$\alpha$ strongly lowers the deflection angle in the outer regions ($\xi >
1$) with only a minor effect for $\xi < 1$. These trends are consistent
with the asymtptotic scalings $\alpha_s \propto \xi^{\alpha - \gamma + 2}$
for $\xi << 1$ and $\alpha_s \propto \xi^{\alpha + \beta -1}$ for $\xi >>
1$. Note that, as expected, for models with finite total mass $(\alpha +
\beta = 0)$, we get $\alpha_s \propto 1/\xi$ as for a point mass lens. Finally,
we note that $\alpha_s$ is an increasing function of $x_0$ which can be
explained by noting that the larger is $x_0$, the more the model is
concentrated and hence more mass is within a circle with radius $\xi$ and
hence the larger is the deflection angle.

\begin{figure*}
\centering \resizebox{15cm}{!}{\includegraphics{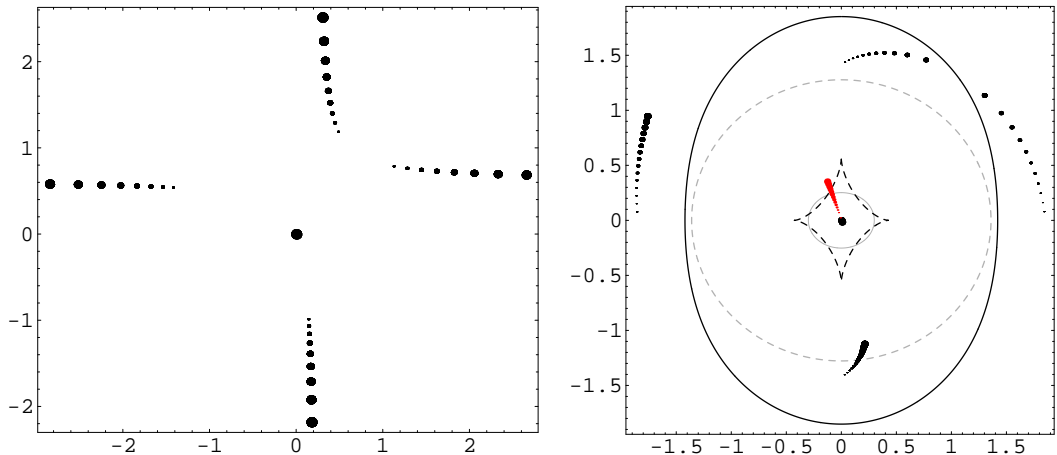}}
\caption{Image configurations for different values of the
parameters\,: the dimension of the spot changes varying the
parameters. {\it Left.} We set the source position $r_s = 0.1''$
and let $\beta$ increasing (the higher is $\beta$, the greater is
the spot).  {\it Right.} Here $\beta$ is fixed at 1 and $r_s$
changes. The bold continue line is the critical curve and the
corresponding caustic is the astroid (dashed bold curve at
center); the internal lighter continue ellipse is the radial
critical curve and the corresponding caustic is the dashed lighter
ellipse. In both the panel we set $\alpha = -0.3$, $\Upsilon_{0} =
5$, $\theta_{s} = 20^{\circ}$, $\gamma_{s} = 0.15$ and
$\theta_{\gamma_{s}} = 180^{\circ}$.} \label{image_positions}
\end{figure*}

The spherical symmetry of the model makes it possible to straightforwardly
compute the lensing potential integrating the deflection angle thus
obtaining\,:

\begin{equation}
\psi = a \int{\hat{\alpha}(\xi) d\xi} = \psi_1 {\times}
\psi_{s}(\xi; {\bf p})
\end{equation}
with $\psi_{s}$ the scaled potential and $\psi_1 = \psi(\xi = 1)$.
The complete expression of $\psi_s$ is too cumbersome and will be
not reported here. However, we plot $\psi_s$ as a function of
$\xi$ for some representative models in the right panels of
Fig.\,\ref{def_angle} showing that $\psi_s$ is an increasing
function of $\beta$. For $\xi \to 0$, $\psi_{s} \propto
\xi^{\alpha - \gamma + 3}$, while $\psi_{s} \propto \xi^{\alpha +
\beta}$ as $\xi \to \infty$. Note that both these trends should be
derived directly from those of $\alpha_s$ given how $\alpha_s$ and
$\psi_s$ are related.

The deflection potential is an essential ingredient to write down
the time delay function and hence estimate the time delay between
the images in the multiply imaged quasars. Adopting polar
coordinates $(R, \theta)$ in the lens plane with $\theta$ measured
counterclockwise from North, it is\,:

\begin{equation}
\Delta t =h^{-1} \tau_{100} \ a^{2} \bigg[ \frac{1}{2}\xi^{2}- \xi
\ \xi_{s} \cos (\theta -\theta_{s}) +
\frac{1}{2}\xi_{s}^{2}-\psi(\xi,\theta) \bigg ],
\end{equation}
where $(\xi, \theta)$ and $(\xi_{s}, \theta_{s})$ are the images
and source positions, $h$ the dimensionless Hubble constant and
$\tau_{100} = (1 + z_d)/c {\times} D_{d}D_{s}/ D_{ds}$ with $z_d$
the lens redshift. According to the Fermat principle, the images
form at the extrema of the time delay function so that the lens
equations are\,:

\begin{equation}
\xi -\xi_{s} \cos(\theta - \theta_{s}) =
\frac{\hat{\alpha}(\xi)}{a}, \label{lens1}
\end{equation}

\begin{equation}
\xi_{s} \sin(\theta -\theta_{s}) = 0 \ . \label{lens2}
\end{equation}
Eq.(\ref{lens2}) reduces to an identity for $\xi_s = 0$, i.e. the
lens and the source are perfectly aligned. In this case, the image
is the well known Einstein ring with radius obtained by solving
Eq.(\ref{lens1}) for $\xi_s = 0$. Alternatively, Eq.(\ref{lens2})
is solved by $\theta = \theta_s + n \pi$ with $n = 0, 1$.
Inserting these values in Eq.(\ref{lens1}) and solving with
respect to $\xi$, we get three images. Two of them are disposed
symmetrically with respect to the lens centre, while the third one
lies very close to the centre (with $\xi \simeq 0$) and turns out
to be unobservable.

A quantitative investigation of the time delay function is shown
in Fig.\,\ref{Dt_all}. In the left panel, we plot $\Delta t$ as
function of the radial distance $r$ for different values of the
model parameters. Note that, as expected, the function $\Delta
t(r)$ has a maximum and two minimum thus meaning that three images
are formed. In the right panels, $r$ is held fixed and we plot
$\Delta t$ as function of $\beta$ for different values of
$\alpha$. An image forming at $r$ will be more and more delayed as
$\alpha$ and $\beta$ get higher. A detailed analysis of the
dependence of $\Delta t$ on the model parameters is quite
complicate because of the many parameters involved (including the
source position) so that will be not performed here.

\subsection{Lens ellipticity and contribution of the shear}

Real galaxies are hardly spherically symmetric. An easy way to
take into account the ellipticity of the lens should be to insert
it {\it by hand} in the lensing potential replacing the
cylindrical dimensionless radius $\xi$ with its elliptical
counterpart $\tilde{m} = \sqrt{\xi_1^2 + \xi_2^2/q^2}$ with $q$
the axial ratio. A still simpler approach is to add an external
shear to the lensing potential since it has been shown that an
elliptical potential $\psi(\tilde{m}^2)$ with an on axis shear
$\gamma_s$ and axial ratio $q$  produce the same image
configuration as a pure elliptical potential with an axial ratio
$q' = q [(1-\gamma_{s})/(1+\gamma_{s})]^{1/2}$ and no shear
(\cite{Witt96}). Moreover, an external shear could also mimic the
effect of tidal perturbations from nearby galaxies or departures
from a smooth distribution because of substructures in the primary
lens. Motivated by these considerations, we add the shear term\,:

\begin{equation}
\psi_{\gamma_{s}} = -\frac{1}{2} \gamma_{s} a^2 \xi^2 \cos{2
(\theta -\theta_{\gamma_{s}})}
\end{equation}
to the lensing potential $\psi$ being $(\gamma_s,
\theta_{\gamma_s})$ the shear strength and direction.

The main effect of the shear is to alter the shape of the critical
curves, defined as the loci in the lens plane where the
magnification formally diverges. In particular, not considering
the radial critical curve, the higher is $\gamma_s$, the larger is
the deformation of the tangential critical curve. In the case of
no shear, this curve coincides with the Einstein ring, while, for
$\gamma_{s} \ne 0$, we have to distinguish two closed curve: the
first one elongated along the direction of the shear itself (the
critical curve) and the other one in perpendicular direction (the
Einstein ring). Moreover, the larger is $\gamma_s$, the lower will
be the axial ratio $b/a$ between the shortest and the longest axis
of these ellipses.

\subsection{Image formation}

As quoted above, the axisymmetric lens can produce only 2 images
with in addition a central (observationally not visible) one. When
we distort the lensing isopotential contours by adding the shear
term, the shape of tangential critical curves changes and the
point\,-\,like caustic unfolds generating four cusps (\cite{SEF}).
It is interesting to discuss the formation of the images as a
function of some of the model parameters for the case with the
shear term included.

In Fig.\,\ref{image_positions}, we plot the image configuration
for some values of the parameters and changing $\beta$ or $r_{s}$.
We can see how the model can generate 3 or 5 images (2 or 4 plus a
central image). In the left panel, we fix the source position and
change the values of $\beta$ finding that the radial coordinate of
the image increases with $\beta$. The right panel shows, for a
fixed $\beta$, how the number and the position of the images
change with the source position $r_s$. Increasing $r_s$ two of the
images approach each other in direction of the tangential critical
curve, until the source exits the region delimited by the
tangential caustic curve (the central astroid) when the two images
merge and the system passes from a 5\,-\,images to a 3\,-\,images
configuration. The other two curves in figure represents the
radial critical curve and the corresponding caustic.

The number of images needs a brief remark. The odd-number theorem
predicts that lenses with a smooth and not singular surface mass
density $\kappa(R)$ which decreases faster than $R^{-1}$ as $\vert
R \vert \to 0$ necessarily forms an odd number of images. For
models with $\kappa$ diverging for $R \to 0$, we can use a
generalized version of the previous theorem given by Evans \&
Wilkinson (1998). These authors consider models with an inner cusp
and $\kappa \propto R^{-n}$ (e.g., NFW models above described) and
distinguish three different behaviours (see Ref. above for major
details)
\begin{enumerate}
\item if $0< n < 1$ (weak density cusp), then we observe an odd number of images
\item if $1 < n < 2$ (strong density cusp), then we observe an even number of images
\item if $n = 1$ (isothermal density cusp), then we can observe both an odd or even number of images
\end{enumerate}
Our model can predict an asymptotically decreasing (flat) $\kappa$
for $\gamma - 2< \alpha <\gamma -1$ ($\gamma - 1< \alpha
<\gamma$).  When $\gamma - 2< \alpha <\gamma -1$ is $n = -(\alpha
- \gamma +1)$ and it is simple to verify that these models belong
to the weak density cusp class, with $0<n<1$, thus an odd number
of images always exist as observed in Fig. \ref{image_positions}
for all the values of $\gamma$. In addition, when $\kappa$ is
cored (for instance, if $\gamma - 1< \alpha <\gamma$), then the
original odd-number theorem holds and predicts also in this case
an odd number of images. Thus, our model cannot reproduce strong
cusp, predicting for all the possible combinations of the lens
parameters an odd number of images and the formation of both
radial and tangential critical curves (the first one not observed
in the case of strong cusp models). Nearly all the observed lenses
present two or four images, thus a discordance appears. However,
observations and theoretical predictions can be reconciled if the
central image is very close to the lens and highly demagnified by
the high central surface density, making this image difficult to
detect. Actually, central images were observed in the 2-image lens
\object{PMN J1632-0033} (Winn et al. 2003, 2004), therefore the
existence of these events can allow to constrain the lens model
and particularly the inner slope of the mass density.

\subsection{Comparison with commonly used lens models}

Having investigated in detail the lensing properties of our
phenomenological model, it is worth comparing it with some of the
most used models in literature.

As a quite general and simple class of models, we consider the
power law lensing potential\,:

\begin{equation}
\psi(R,\theta) = R^{\nu} F(\theta, q,
\theta_{q})\label{eq:ell_pot}
\end{equation}
where $0 < \nu < 2$ is the slope of the radial profile and
$F(\theta, q, \theta_q)$ the angular shape function with $(q,
\theta_q)$ the axial ratio and the orientation of the isopotential
contours. Eq.(\ref{eq:ell_pot}) represents a generalization of the
the pseudo\,-\,isothermal elliptical potentials (\cite{K87,KK95})
that have been designed to be as similar as possible to the
pseudo\,-\,isothermal mass distribution models (\cite{BK87,KK95}).
Given their easy manageability, they have been widely used in
literature. For semianalytical and/or numerical treatments in
modeling lensed quasars see, e.g., Cardone et al. (2001, 2002),
Tortora et al. (2004). In addition, considering that $\kappa
\propto R^{\nu - 2}$ and following the previous Sec., power law
with $0<\nu< 1$ ($1<\nu <2$) have strong (weak) cusp.

It is worth stressing that the models defined by
Eq.(\ref{eq:ell_pot}) are characterized by a constant radial slope
of the lensing potential.  Obviously, this is not the case for our
phenomenological model. Nevertheless, we can match the two models
in the inner regions by setting $\alpha - \gamma + 3 = \nu$. In
particular, for $\gamma = 1$ and $\xi \rightarrow 0$, our model
reproduces the pseudo\,-\,isothermal trend for $\alpha \in [-1,0]$
thus excluding the region $(0,1]$, since for these values of
$\alpha$ is $\nu > 2$ and the pseudo\,-\,isothermal models
generate a non physical surface mass density. It is possible to
match the outer slope imposing $\alpha + \beta = \nu$: also in
this case the matching is not complete since when $\alpha + \beta
>1$ our models are not defined, on the contrary the models in Eq.
(\ref{eq:ell_pot}) holds also for $\nu \in (1,2]$\footnote{ This
circumstance is verified since we required that our model never
has an increasing $v_{c}$ (giving the constraint $\alpha + \beta
<1$), on the contrary models (\ref{eq:ell_pot}) predict both
decreasing and increasing $v_{c}$.  }.

\section{Conclusions}

The rotation curves of spiral galaxies have been the first
observational cornerstone of the dark matter theory. The nature
and the distribution of this unseen component of galaxies are
still obscure, but more and more evidences have been accumulated
in favour of its ubiquitous presence. Nonetheless, this picture is
far to be free of its own problems since recent observations of
line of sight velocity profile of elliptical galaxies have
furnished contrasting evidences favoring or disfavoring the
presence of significant amounts of dark matter in the outer
regions of these systems. In an attempt to face this problem we
have employed here a phenomenological approach assuming a double
power law expression for the global $M/L$ ratio as $\Upsilon(r) =
\Upsilon_0 (r/r_0)^{\alpha} (1 + r/r_0)^{\beta}$.  This simple
parametrization reduces to constant $M/L$ models for $(\alpha,
\beta) = (0, 0)$, while gives an asymptotically increasing $M/L$
ratio for $\alpha + \beta \ge 0$ as in models with dominating dark
matter component. Adopting spherical symmetry and a versatile
expression for the luminosity density $j(r)$, it is possible to
build a wide class of (effective) galaxy models starting from the
mass profile $M(r) = \Upsilon(r) L(r)$. Varying the model
parameters $(\alpha, \beta)$ in the range determined by some
physical considerations makes it possible to mimic a wide range of
models having finite $(\alpha + \beta = 0)$ or formally infinite
$(\alpha + \beta > 0)$ total mass, cuspy $(\alpha - \gamma < 0)$
or cored $(\alpha - \gamma = 0)$ density profiles, asymptotically
flat $(\alpha + \beta - 1 = 0)$ or decreasing $(\alpha + \beta - 1
< 0)$ rotation curve, cuspy $(\gamma -2 <\alpha < \gamma -1)$ or
cored $(\gamma -1 <\alpha < \gamma)$ surface mass density.

The main advantage of our parametrization of the $M/L$ model is
therefore the possibility to reproduce a wide phenomenology
resorting to a simple analytical expression and relying on
phenomenological considerations only thus avoiding any theoretical
prejudice or systematic errors related to convergence problems in
numerical N\,-\,body simulations of structure formation. As a
preliminary and mandatory step, to allow a future comparison
between theoretically predicted and observed properties of early
type galaxies, we have here derived the main dynamical and lensing
quantities of the model giving whenever possible analytical
expressions or illustrative plots. We have payed a particular
attention to the asymptotic trends in the inner and outer regions
since both these regimes are mostly probed by different kind of
observations. On the one hand, measurement of the velocity
dispersion profiles hardly extends to radii larger than the
effective radius $R_e$ and, indeed, only the central velocity
dispersion is often at disposal. Such a mass tracer therefore
mainly probes the inner region of the system so that knowing how
our model fares for $x << 1$ helps in comparing with this kind of
data. In order to probe larger radii, one may resort to multiply
imaged lensed quasars. First, the images typically form at $r \sim
(1\,-\,2) R_e$ so that a larger region is probed. Moreover, the
lensing potential and the flux ratios also depend on the
derivatives of the mass profile thus allowing to probe indirectly
also outermost regions.

We can individuate two main implications. Firstly, we have seen
how the the full mass density profile and derived DM profile
depend on $\gamma$ and therefore on the fitted luminous profile.
Therefore, the presence of the baryons influences dark matter
profile, since the parameter $\gamma$ that specifies the inner
slope of the luminous profile enters in the expression of halo
profile derived from our model. Secondly, we have compared our
model with the more common generalized NFW profile: the asymptotic
trends has been compared and we are able with suitable choices of
$\alpha$ and $\beta$ to recover the behaviour of gNFW model. The
value of the inner trend of mass profile has generated a long
debate in literature. The simulations seems to indicate an inner
trend that scales like $r^{-\delta}$ with $\delta \geq 1$ (the so
called cusped models). On the contrary, many results from
observational analysis, by rotation curves of low surface
brightness, dwarf galaxies and spiral galaxies, dynamical studies
of elliptical galaxies and gravitational lensing
(\cite{MdeB98,BE01,BS2001,deB01,TK02b,BSD2003}) suggested the
presence of internal core with $\delta < 1$ (cored models). This
contradiction generates the so called cusp/core problem. Our model
can reproduce both cuspy models with $\delta \geq 1$ and cored
ones with $\delta < 1$. In details, $\alpha < 0$ recovers the
first models, i.e. the results from simulations and $\alpha \geq
0$ the seconde one, those from results of observations. Being able
to reproduce all these different behaviours, our phenomenological
model offers the perspective to homogenously analyze the
observations and give an answer to the cusp/core problem.

On a theoretical side, our approach could be further ameliorated.
A key ingredient in our modeling is the luminosity density $j(r)$
which we have assumed to be described by the class of Dehnen
models. Although this gives a quite general expression able to
reproduce well the observed surface brightness profile, the ideal
route to follow should be to deproject an empirical function
obtained by a direct fit to the photometric data. However, this
approach should be repeated for each single galaxy and thus does
not allow to investigate general properties. An intermediate
possibility could be to use the deprojection of the Sersic profile
as input for the luminosity density and investigate the dynamical
and lensing properties of the corresponding model. Moreover, it
has been shown that the dynamics of inner regions of elliptical
galaxies is better described adding the contribution of the
central supermassive black hole
(\cite{BD04,MamonLokas05a,MamonLokas05b}) so that it is
interesting to update our parametrization including this term into
the mass profile. Finally, departures from spherical symmetry
could also be investigated since real galaxies are likely to be
moderately triaxial systems (\cite{FGdeZ}).

Contrasting the model with the observations is, of course, the
best strategy to both test the viability of our parametrization
and constrain its parameters. Given the difficulties in obtaining
radial velocity dispersion profile up to large radii, an ideal
tool could be represented by lens galaxies. With more than 90
observed multiply imaged quasars, gravitational lensing offers a
unique dataset to put severe constraints on the slope parameters
$(\alpha, \beta)$ and hence suggest useful hints to solve the
problem of dark matter in elliptical galaxies. Moreover, since
lens galaxies span a range in redshift from low to intermediate,
such an analysis also offers the possibility to investigate
whether and how the characteristics of the dark matter content of
early type galaxies evolve.

A first qualitative result can be obtained considering the index
of the surface mass density defined as ${\rm ind}{\kappa(r)} =
d\ln{\kappa}/d\ln{r}$. In order to shape lensing events with their
model independent analysis, Williams \& Saha (2000) have used as a
constraint on the chosen lens model the condition ${\rm
ind}{\kappa(r)} \le -0.5$ for values of $r$ close to the images
positions in strongly lensed quasars (see also \cite{Tortora}).
Such a condition ensures that the image magnification is less than
$\sim 2$ which is a quite plausible constraint well satisfied in
some real lenses. Considering that typically images forms at
$(1\,-\,2)R_e$, we have checked that the constraints on ${\rm
ind}{\kappa(r)}$ is satisfied by our model for all physically
reasonable choices of the slope parameters. Although this does not
allow to further constrain $(\alpha, \beta)$, this result suggests
that our parametrization of the $M/L$ ratio leads to a model that
correctly reproduces this aspect of the observed lensing
phenomenology.

A detailed investigation of some interesting cases is outside our
aim here and will be presented in a forthcoming paper. Here, we
only note that, although the luminous component parameters
$(\gamma, a)$ may be set from the analysis of the surface
brightness profile, severe degeneracies still exist among the
slope parameters $(\alpha, \beta)$ and the scaling quantities
$(\Upsilon_0, x_0)$. As a consequence, it is likely that reliable
constraints could be obtained only considering bright quadruply
imaged systems so that the errors on the image positions is low
and the number of constraints higher. Moreover, it is preferable
to limit the attention to isolated lens galaxies in order to avoid
to introduce further unknown parameters to describe the lens
environment. Finally, it is highly desirable to have a measurement
of the central velocity dispersion in order to break some of the
degeneracies inherent in lens modeling with the help of dynamical
data (\cite{TK02,TK04}).

As a final comment, we would like to stress again the need for a
phenomenological approach to the problem of dark matter in
elliptical galaxies. In order to overcome any theoretical
prejudice, one should adopt a procedure that matches a large set
of observational constraints with the minimal number of
hypotheses. Adding gravitational lensing data to the dynamical one
makes it possible to advance our knowledge on the observational
side of the problem. Using phenomenological models like the one we
have presented here could be the theoretical ingredient best
suited to be coupled to the above data to analyze the question of
dark matter in early-type galaxies.

\begin{acknowledgements}
We warmly thank the anonymous referee for his report that have
helped us to significantly improve the paper.
\end{acknowledgements}


\begin{thebibliography}{99}

\bibitem[Baes \& Dejonghe 2004]{BD04}
Baes, M., Dejonghe, H. 2004 MNRAS, 351, 18

\bibitem[Bertin et al. 1994]{B94}
Bertin, G. et al. 1994, A\&A, 292, 381

\bibitem[Binney \& Tremaine 1987]{BT87}
Binney, J., Tremaine, S. 1987, {\it Galactic dynamics}, Princeton
University Press

\bibitem[Binney \& Evans 2001]{BE01}
Binney, J. J. \& Evans, N. W. 2001, MNRAS, 327, L27

\bibitem[Blandford \& Kochanek 1987]{BK87}
Blandford, R., Kochanek, C.S. 1987, ApJ, 321, 658

\bibitem[Borriello \& Salucci 2001]{BS2001}
Borriello, A., Salucci, P., 2001, MNRAS, 323, 285

\bibitem[Borriello et al. 2003]{BSD2003}
Borriello, A., Salucci, P., Danese, L. 2003, MNRAS, 341, 1109

\bibitem[Caon et al. 1993]{CCD93}
Caon, N., Capaccioli, M., D' Onofrio, M. 1993, MNRAS, 265, 1013

\bibitem[Cardone et al. 2001]{PULP}
Cardone, V.F., Capozziello, S., Re, V., Piedipalumbo, E. 2001
A\&A, 379, 72

\bibitem[Cardone et al. 2002]{HERQULES}
Cardone, V.F., Capozziello, S., Re, V., Piedipalumbo, E. 2002
A\&A, 382, 792

\bibitem[Carroll et al. 2005]{carroll} Carroll, S.M et al. 2005, PRD 71, 063513

\bibitem[de Blok et al. 2001]{deB01}
de Blok, W. J. G., McGaugh, Stacy S., \& Rubin, V. C. 2001, AJ,
122, 2396

\bibitem[Dehnen 1993]{Dehnen93}
Dehnen, W. 1993, MNRAS, 265, 250

\bibitem[de Vaucouleurs 1948]{deV48}
de Vaucouleurs, G. 1948, Ann. d' Ap., 11, 247

\bibitem[Evans \& Wilkinson 1998]{EW98}
Evans, N. W. \& Wilkinson, M. I. 1998, MNRAS, 296, 800

\bibitem[Fisher et al. 2000]{Fisher00}
Fisher, P., et al., 2000, AJ, 120, 1198

\bibitem[Franx et al. 1994]{FGdeZ}
Franx, M., van Gorkom, J.H., de Zeeuw, P.T. 1994, ApJ, 436, 642


\bibitem[Fukugita \& Turner 1991]{FT91}
Fukugita, M., Turner, E.L. 1991, MNRAS, 253, 99

\bibitem[Fukushige \& Makino 2001]{FM01}
Fukushige, T. \& Makino, J. 2001, ApJ, 557, 533

\bibitem[Gerhard et al. 2001]{Ger01}
Gerhard, O., Kronawitter, A., Saglia, R.P., Bender, R. 2001, AJ,
121, 1936
\bibitem[Ghigna et al. 2000]{Ghigna2000}
Ghigna, S., Moore, B., Governato, F., et al. 2000, ApJ, 544, 616

\bibitem[Gradshteyn \& Ryzhik 1980]{GR80}
Gradshteyn, I.S., Ryzhik, I.M. 1980, {\it Table of integrals,
series and products}, Academic Press

\bibitem[Graham \& Colless 1997]{GC97}
Graham, A., Colless, M. 1997, MNRAS, 287, 221

\bibitem[Guzik \& Seljak 2002]{GS02}
Guzik, J., Seljak, U. 2002 MNRAS, 335, 311

\bibitem[Hawkins et al. 2003]{H03}
Hawkins, E. et al. 2003, MNRAS, 346, 78

\bibitem[Hernquist 1990]{H90}
Hernquist, L. 1990, ApJ, 356, 359

\bibitem[Jaffe 1983]{J83}
Jaffe, W. 1983, MNRAS, 202, 995

\bibitem[Jing \& Suto 2000]{JS00}
Jing, Y. P. \& Suto, Y. 2000, ApJ, 529, L69


\bibitem[Kassiola \& Kovner 1995]{KK95}
Kassiola, A., Kovner, I. 1995, MNRAS, 272, 363

\bibitem[Keeton et al. 1998]{KKF98}
Keeton, C.R., Kochanek, C.S., Falco, E.E. 1998, ApJ, 509, 561

\bibitem[Kochanek 1995]{K95}
Kochanek, C.S. 1995, ApJ, 445, 559

\bibitem[Kochanek 2002]{Koch02}
Kochanek, C.S. 2002, ApJ, 578, 25

\bibitem[Kovner 1987]{K87}
Kovner, I. 1987, Nat, 325, 507

\bibitem[Loewenstein \& White 1999]{LW99}
Loewenstein, M., White, R.E. 1999 ApJ, 518, 50

\bibitem[Macci{\`o} et al. 2006a]{M2006a}
Macci{\`o}, A.V., Moore, B., Stadel, J. and Diemand, J. MNRAS,
2006a, 366, 1529

\bibitem[Macci{\`o} et al. 2006b]{M2006b}
Macci{\`o}, A.V., Moore, B. and Stadel, J. ApJL, 2006b, 636, L25


\bibitem[Magorrian \& Ballantyne 2001]{MB01}
Magorrian, J., Ballantyne, D. 2001, MNRAS, 322, 702

\bibitem[Mamon \& Lokas 2005a]{MamonLokas05a}
Mamon, G.A., Lokas, E.L. 2005a, MNRAS, 362, 95

\bibitem[Mamon \& Lokas 2005b]{MamonLokas05b}
Mamon, G.A., Lokas, E.L. 2005b, MNRAS, 363, 705

\bibitem[Mazure \& Capelato 2002]{MC02}
Mazure, A., Capelato, H.V. 2002, A\&A, 383, 384

\bibitem[McGaugh \& de Blok 1998]{MdeB98}
McGaugh, S. S. \& de Blok, W. J. G. 1998, ApJ, 499, 66

\bibitem[Mould et al. 1990]{M90}
Mould, J.R., Oke, J.B, de Zeeuw, P.T., Nemec, J.M. 1990, AJ, 99,
1823

\bibitem[Merritt 1985]{Merritt85}
Merritt, D. 1985, MNRAS, 214, 25

\bibitem[Moore et al. 1998]{M98}
Moore, B., Governato, F., Qinn, T., Stadel, J., Lake, G. 1998,
ApJ, 499, L5

\bibitem[M$\ddot{u}$cket \& Hoeft 2003]{MH03}
M$\ddot{u}$cket, J. P. \& Hoeft, M. 2003, A\&A, 404, 809.

\bibitem[Napolitano et al. 2004]{Nap04}
Napolitano, N.R. et al. 2004, astro\,-\,ph/0411639

\bibitem[Navarro et al. 1996]{NFW96}
Navarro, J.F., Frenk, C.S., White, S.D.M. 1996, ApJ, 462, 563

\bibitem[Navarro et al. 1997]{NFW97}
Navarro, J.F., Frenk, C.S., White, S.D.M. 1997, ApJ, 490, 493

\bibitem[Navarro et al. 2004]{N03}
Navarro, J.F., Hayashi, E., Power, C., Jenkins, A.R.,  Frenck,
C.S. 2004, MNRAS, 349, 1039

\bibitem[Osipkov 1979]{Osipkov79}
Osipkov, L.P. 1979, Soviet Astronomy Letters, 5, 42

\bibitem[Petters et al. 2001]{PLW01}
Petters, A.O., Levine, H., Wambsganss, J. 2001, {\it Singularity
theory and gravitational lensing}, Birkh\"{a}user, Boston

\bibitem[Pope et al. 2004]{P04}
Pope, A.C. et al. 2004, ApJ, 607, 655

\bibitem[Power et al. 2003]{P03}
Power, C., Navarro, J.F., Jenkins, A., Frenk, C.S., White, S.D.M.,
Springel, V., Stadel, J., Quinn, T. 2003, MNRAS, 338, 14

\bibitem[Prugniel \& Simien 1997]{PS97}
Prugniel, Ph., Simien, F. 1997, A\&A, 321, 111

\bibitem[Refsdal 1964]{Refsdal}
Refsdal, S. 1964, MNRAS, 128, 307

\bibitem[Riess et al. 2004 ]{Riess04}
Riess A.G., et al., ApJ, {\bf 607}, 665

\bibitem[Romanowsky \& Kockanek 1997]{RK97}
Romanowsky, A.J., Kochanek, C.S. 1997, MNRAS, 287, 95

\bibitem[Romanowsky et al. 2001]{rom01}
Romanowsky, A.J. et al. 2001, ApJ, 553, 722

\bibitem[Romanowsky et al. 2003]{Romanowsky03}
Romanowsky, A.J. et al. 2003, Science, 301, 1696

\bibitem[Sand et al. 2004]{Sand04}
Sand, D. J., Treu, T., Smith, G.P., Ellis, R.S. 2004, ApJ, 604, 88


\bibitem[Schneider, Ehlers \&  Falco 1992]{SEF}
Schneider, P., Ehlers, J.,  Falco, E.E. 1992, {\it Gravitational
lenses}, Springer\,-\,Verlag, Berlin

\bibitem[Sersic 1968]{Sersic}
Sersic, J.L. 1968, {\it Atlas de Galaxies Australes}, Observatorio
Astronomico de Cordoba


\bibitem[Sofue \& Rubin 2001]{SR01}
Sofue, Y., Rubin, V. 2001, ARA\&A, 39, 137


\bibitem[Spergel et al. 2003]{WMAP}
Spergel, D.N. et al. ApJS, 148, 175, 2003


\bibitem[Tortora et al.  2004]{Tortora}
Tortora, C., Piedipalumbo, E., Cardone, V.F. 2004, MNRAS, 354, 343

\bibitem[Treu \& Koopmans 2002a]{TK02}
Treu, T. \& Koopmans, L. V. E. 2002a, MNRAS, 337, L6

\bibitem[Treu \& Koopmans 2002b]{TK02b}
Treu, T. \& Koopmans, L. V. E. 2002b, ApJ, 575, 87

\bibitem[Treu \& Koopmans 2004]{TK04}
Treu, T., Koopmans, L.V.E. 2004, ApJ, 611, 739

\bibitem[Turner et al. 1984]{TOG84}
Turner, E.L., Ostriker, J.P., Gott, J.R. 1984, ApJ, 284, 1


\bibitem[Williams \& Saha 2000]{WS2000}
Williams, L.L.R., Saha, P. 2000, AJ, 119, 439

\bibitem[Winn et al. 2003]{WRK03}
Winn J. N., Rusin D., Kochanek C. S., 2003, ApJ, 587, 80

\bibitem[Winn et al. 2004]{WRK04}
Winn J. N., Rusin D., Kochanek C. S., 2004, Nature, 427, 613

\bibitem[Witt 1996]{Witt96}
Witt, H.J. 1996, ApJ, 472, L1

\bibitem[Zwicky 1937]{Zwicky}
Zwicky, F. 1937, ApJ, 86, 217

\end{thebibliography}
\end{document}